\documentclass[twocolumn,showpacs,preprintnumbers,amsmath,amssymb]{revtex4-2} 
\usepackage{graphicx}
\usepackage{dcolumn}
\usepackage{bm}
\usepackage{tabularx}
\usepackage{ulem} 
\newcolumntype{M}{>{\centering\arraybackslash}m{1.85cm}}
\usepackage[export]{adjustbox}
\usepackage{float}
\usepackage{braket}
\usepackage{lipsum}
\usepackage{amsmath}
\graphicspath{{figure/}}
\usepackage{xcolor}   

\usepackage{caption}
\usepackage{subcaption}

\usepackage{hyperref}
\hypersetup{
    colorlinks=true, 
    linktoc=all,     
    urlcolor  = cyan,
    citecolor = blue,
    linkcolor=blue,  
}

\makeatletter
\newcommand{\colorcaption}[2][]{%
  \begingroup%
  \renewcommand{\@caption@fignum@sep}{ (Color online). }%
  \caption[#1]{#2}%
  \endgroup%
}
\makeatother
\newcommand{\orcid}[1]{\href{https://orcid.org/#1}{\hskip2pt\includegraphics[width=9pt]{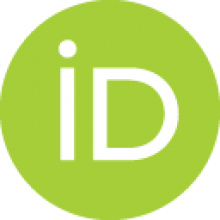}}}
\begin{document}
\title{Entanglement study in the island of inversion region using \textit{ab initio} approach}

\author{Rohit M. Shinde\orcid{0009-0007-0933-038X}}
\email{rohit\_ms@ph.iitr.ac.in}
\affiliation{Department of Physics, Indian Institute of Technology Roorkee, Roorkee 247667, India}

\author{Praveen C. Srivastava\orcid{0000-0001-8719-1548}}
\email{ praveen.srivastava@ph.iitr.ac.in}
\affiliation{Department of Physics, Indian Institute of Technology Roorkee, Roorkee 247667, India}

\date{\hfill \today}

\begin{abstract}
Quantum entanglement provides a unique perspective for probing nuclear structure. In this work, we employ quantum entanglement measures, including proton-neutron entanglement entropy, mutual information, and quantum relative entropy, to investigate the evolution of entanglement patterns as we approach neutron-rich nuclei. The study is carried out in the vicinity of the $N=20$ island of inversion region consisting of even-$A$ Ne, Mg, and Si isotopes, and also for isotones corresponding to $N=20$. The state-of-the-art \textit{ab initio} valence space in-medium similarity renormalization group method has been used for this purpose. We have highlighted the role of proton-neutron entanglement entropy in the formation of the island of inversion region.  
Mutual information provides insight into the strength of correlations between proton-proton, neutron-neutron, and proton-neutron single-particle states. While these correlations are relatively weak between protons and neutrons in the ground states, they become comparable to like-particle correlations in excited states. The quantum relative entropy is also studied between $0^+$ and $2^+$ states of the Ne, Mg, and Si isotopes, as well as $N=20$ isotones, using the Kullback-Leibler divergence and Jensen-Shannon divergence. 
 We have performed these calculations by expressing the nuclear wavefunctions in a Slater-determinant basis and analyzing them through complementary partitions, including proton-neutron and mode-resolved factorizations.

\end{abstract}


\maketitle

\section{INTRODUCTION}
Over the past two decades, advances in quantum computing algorithms and quantum simulation platforms have significantly expanded the scope of quantum information science \cite{chuang,nisq,vqe,Perez2_2023,Ayral_Besserve_Lacroix_2023,Lacroix_PRL,Chandan_2023,Bharti_Bhoy,QC_n1,QC_n2,graycode,pooja2,QC_n3,QC_n4,QC_n5,Robin_Savage_2023,Denis_2024,Yoshida_2024,Dean,Vary_2021}. These developments have established entanglement as a reliable measure to quantify correlations in quantum many-body systems
\cite{Horodecki_2009,Hill_Wootters_1997,Belavkin_Ohya_2002,Eisert_2006,Gigena_Rossignoli_2015}. The study of entanglement in fermionic systems is a challenging task due to the indistinguishability of the particles \cite{Horodecki_2009,Hill_Wootters_1997,Belavkin_Ohya_2002,Eisert_2006,Gigena_Rossignoli_2015,Schliemann_2001,Eckert_2002,Wiseman_Vaccaro_2003,Ghirardi_Marinatto_2004,Benatti_Floreanini_Titimbo_2014,LoFranco_Compagno_2016}. The explanation lies in the fact that, for indistinguishable particles, the many-body Hilbert space does not possess a straightforward tensor-product framework, complicating the partitioning of the system. This challenge is resolved through the use of the Fock space representation, in which states are articulated using occupation numbers \cite{Benatti_Floreanini_Marzolino_2014,Zanardi_2002,Shi_2003,Friis_2013}. This framework enables the Hilbert space to be partitioned into bipartite and multipartite subsystems, and subsequently calculate corresponding entanglement measures \cite{Coffman_Kundu_Wootters_2000,Wong_Christensen_2001}.    

The atomic nucleus is a finite, strongly correlated quantum many-body system, making it a suitable candidate for studying the role of entanglement and exploring nucleon-nucleon correlations from a new perspective. Several attempts have already been made to understand the role of entanglement in nuclear systems using various models such as the Lipkin-Meshkov-Glick model \cite{Baid_2024,Hengstenberg_2023,Latorre_2005,Vidal_Palacios_Aslangul_2004}, the no-core shell model \cite{Carl_2021,Sarma_2024}, the SEN model \cite{Kruppa_2022}, density matrix renormalization group (DMRG) approach \cite{Tichai_2023,Pazy_2023}, pairing models \cite{Liang_2025,Gu_2023}, nucleon-nucleon scattering \cite{Bai_Ren_2022,Bai_2023}, and the spherical shell model \cite{Carl_2025,CJ_2023,Perez_2023,Kruppa_2021,Kruppa_2025,PerezObiol_2024,Shinde_2025,Costa_2025}. There have also been studies that have incorporated the entanglement entropy to improve the nuclear theory calculations \cite{Johnson_2025,CJ_2024}. Quantum information measures studied from a nuclear physics perspective include entanglement entropy, $n$-tangles, negativity, mutual information, and concurrence.

Apart from calculating quantum information measures within the framework of the same nucleus, relative entropy and its quantum extension allow us to study the evolution of structural changes between two different states of the same nucleus. Relative entropy, also known as Kullback-Leibler divergence (KLD), is a fundamental concept in information theory, with applications in machine learning and to measure deviation from thermodynamic equilibrium in statistical mechanics \cite{Vedral_2002,Schumacher_Westmoreland_2000,Sagawa_2012,Floerchinger_2020}. Quantum extensions of KLD have been developed to quantify the distinguishability between quantum states. In addition to the KLD, the Jensen–Shannon divergence (JSD) provides a symmetrized and bounded measure of distinguishability between probability distributions and quantum states. Unlike the KLD, the JSD is finite and well defined, and its square root defines a true metric \cite{Majtey_2005,Grosse_2002,Lamberti_2008,Briet_2009}. 

Although the nuclear shell model has successfully described magic numbers, the exploration of neutron-rich nuclei has revealed departures from standard shell closures \cite{Caurier_2005,BrownReview,Otsuka_2020,SorlinReview}. This phenomenon is observed for neutron-rich $Z=10-12$ isotopes, where the conventional $N=20$ magic number vanishes. This region is referred to as the ``island of inversion" (IoI), where the low-lying states become dominated by intruder configurations involving cross-shell excitations from the $sd$ to the $pf$ shell \cite{WarburtonIoI,Poves1994}. Several studies have employed phenomenological interactions \cite{Kaneko_2011,MCSM_1999,SDPF-U,SDPF-MU,SDPF-U-MIX}, as well as the extended Kuo-Krenciglowa (EKK) method, within many-body perturbation theory \cite{EEdf1}. However, an accurate description of the IoI typically requires either fitting matrix elements to experimental data or empirically adjusting the single-particle energies, respectively. 

To achieve a fully microscopic description of the IoI, \textit{ab initio} approaches based on realistic nuclear forces are required \cite{RagnarReview,HergertReview}. In recent years, the in-medium similarity renormalization group (IMSRG) has emerged as a powerful framework for deriving an effective Hamiltonian directly from chiral two- and three-nucleon interactions \cite{Li2023PLB,IMSRG_Magnus,IMSRG3f2,IMSRG3N7_Heinz2021,IMSRG3N7_Ragnar,IMSRG3N7_Heinz2024}. Its valence-space formulation (VS-IMSRG) enables the decoupling of a chosen valence space from the full Hilbert space \cite{IMSRG_RagnarPRL,IMSRG_RagnarPRC,Miyagi2022PRC}. Recent developments have extended the VS-IMSRG to multishell valence spaces, making it possible to address the $sd$–$pf$ model space relevant for the IoI without empirical adjustments to single-particle energies or interaction matrix elements \cite{IMSRG_Miyagi,Yuan_Hu_2024,Subhrajit1,Subhrajit2}. These advances have demonstrated that the collapse of the $N=20$ shell gap and the emergence of intruder-dominated configurations can be directly described in terms of the underlying nuclear forces. 

In this work, we have investigated the quantum information measures for the shell model eigenstates obtained from VS-IMSRG  calculations in the $sdpf$ model space. We investigate the proton-neutron entanglement entropy for the isotopic chains ${}^{22-34}$Ne, ${}^{24-36}$Mg, and ${}^{26-38}$Si for the first $0^+$, $2^+$, and $4^+$ states of these nuclei, along with the ground states of $N$=20 isotones from ${}^{29}$F to ${}^{35}$P, providing a systematic study of quantum correlations across and beyond the IoI region. As for the mutual information, we have considered the isotopic chains ${}^{24-34}$Ne, ${}^{26-36}$Mg, and ${}^{30-38}$Si and their first $0^+$ and $2^+$ for our study. We have computed the classical KLD, JSD, and their quantum extensions for the same isotopic chains studied in the mutual information analysis, as well as for the $N=20$ isotones, comparing the first $0^+$ and $2^+$ states in even-even nuclei and the ground and first excited states in odd-even nuclei.

This paper is organized as follows. In Section \ref{IMSRG}, we present the framework for \textit{ab initio} calculations using the IMSRG approach, after which we have discussed the representation of many-body wavefunction in Fock space for clean partitioning of the total wavefunction in Section \ref{many-body}. In Sections \ref{proton-neutron}, \ref{MI}, and \ref{QRE}, we have introduced the framework for various quantum information measures. In Section \ref{results}, we present our results corresponding to the aforementioned measures, and finally conclude in Section \ref{summary}.
\section{THEORETICAL FRAMEWORK} \label{framework}

\subsection{ \textit{Ab Initio} Hamiltonian} \label{IMSRG}

In the present work, we have constructed an effective interaction using the IMSRG approach \cite{IMSRG_RagnarPRL,IMSRG_RagnarPRC}.
We start from the intrinsic $A$-body Hamiltonian given by
\begin{equation} \label{eq1}
    H = \frac{1}{A}\sum_{i<j}^A  \frac{{(\vec{p_i}-\vec{p_j})}^2}{2m} +
         \sum_{i<j}^A V_{ij}^{NN} + \sum_{i<j<k}^A V_{ijk}^{NNN},
\end{equation}
where $m$ represents the nucleon mass, $\vec{p}$ is nucleon momentum in laboratory frame. The $V_{ij}^{NN}$ and $V_{ijk}^{NNN}$ correspond to $NN$ and $NNN$ nuclear forces, respectively. The well-established chiral $2N $and $3N$ EM1.8/2.0 interaction \cite{Simonis_2016,Hebeler_2011}, comprising of a next-to-next-to-next-to-leading order (N$^3$LO) $NN$ potential evolved through Similarity Renormalization Group (SRG) to momentum resolution scale $\lambda$=1.8 fm$^{-1}$ and a $NNN$ force at next-to-next-to-leading order (N$^2$LO) with momentum cutoﬀ $\Lambda$=2.0 fm$^{-1}$, has been used for this purpose.

We have performed the IMSRG calculations in a harmonic oscillator basis at $\hbar \omega$=16 MeV with $e=2n+l \leq e_{max}= 12$, and additional truncation on $NNN$ forces $e_1+e_2+e_3 \leq E_{3max}=24$, large enough to reach convergence \cite{Miyagi2022PRC}. The VS-IMSRG interactions are decoupled for the $sd$-shell for both protons and neutrons above $^{16}$O core. The valence-space effective Hamiltonians are diagonalized, and the wavefunction coefficients are extracted through the BIGSTICK code \cite{Johnson_2018}. Recently, our group has reported the nuclear structure properties of the lighter mass region based on \textit{ab initio} theory in Refs.
\cite{Subhrajit1,Subhrajit2,Praveen,Na_work_NPA,Priyanka1,Priyanka2,Chandan1}. The present work is a further extension to apply \textit{ab initio} theory to quantum information. In Section \ref{results}, we have reported all results corresponding to VS-IMSRG interactions.

\subsection{Construction of many-body wavefunction} \label{many-body}
The eigenstates $|\psi\rangle$ obtained by diagonalizing the many-body Schr\"odinger equation are expanded in a complete set of antisymmetric many-body basis states,
\begin{equation}
    |\psi\rangle = \sum_{\alpha} c_{\alpha} |\phi_{\alpha}\rangle .
\end{equation}
Here, $\{|\phi_{\alpha}\rangle\}$ denotes Slater determinants constructed from a chosen single-particle basis, and $c_{\alpha}$ are the corresponding expansion coefficients.

Since we are dealing with a system of fermions, the total wavefunction $|\psi\rangle$ must be antisymmetric under the exchange of particles. A convenient way to represent such many-body basis states is through Slater determinants, which are anti-symmetrized products of single-particle wavefunctions. The single-particle basis is denoted by their quantum numbers $\{i\}=\{n_i,l_i,j_i,m_i,\tau_i\}$, where $n_i$ is the principal quantum number, $l_i$ and $j_i$ are orbital and total angular momenta, $m_i$ is the total angular momenta projection, and $\tau_i$ is the isospin projection. Now the many-body basis states can be constructed by acting with creation operators $\hat{a}_i^\dagger$ on the fermionic vacuum state $|0\rangle$,
\begin{equation}
    |\phi_{\alpha}\rangle = \prod_{i\in\alpha} \hat{a}_i^\dagger |0\rangle .
\end{equation}
Each Slater determinant $|\phi_{\alpha}\rangle$ corresponds to a specific occupation pattern of the single-particle orbitals.

In this work, we adopt the occupation representation of Slater determinants \cite{Suhonen,Johnson_2013}. Using this formalism, the many-body basis state can be represented as
\begin{equation}
 |\phi_{\alpha}\rangle=|n^{\alpha}_1n^{\alpha}_2...n^{\alpha}_N\rangle,   
\end{equation}
where $n^{\alpha}_i$=0 (single-particle state unoccupied), 1 (single-particle state occupied). Finally, the eigenstate can be written as  
\begin{equation}
\begin{aligned}
|\psi\rangle \;&=\; \sum_{\alpha} c_{\alpha}
|n^{\alpha}_1 n^{\alpha}_2 \ldots n^{\alpha}_N\rangle \\
\;&\equiv\; \sum_{n_1,n_2,\ldots,n_N}
c(n_1,n_2,\ldots,n_N)
|n_1 n_2 \ldots n_N\rangle .
\end{aligned}
\label{wavefn}
\end{equation}

Where each set $(n_1,n_2,\ldots,n_N)$ uniquely defines a many-body configuration. The calculations are carried out in the $M$-scheme, with all basis states constructed to have a fixed total angular momentum projection.

\subsection{Proton-neutron entanglement entropy} \label{proton-neutron}
Different ways of partitioning the nuclear wavefunction can reveal valuable information about the nature of entanglement among the various components of the nucleus. A particularly natural choice is to treat protons and neutrons as two distinct subsystems, forming an intrinsic bipartition of the nuclear system  \cite{CJ_2023}. For simplicity, we introduce compact notations for the occupation-number configurations in each subspace:
\begin{equation}
    n_{\pi}\equiv(n^{\pi}_1 n^{\pi}_2..n^{\pi}_{N_{\pi}}),n_{\nu}\equiv(n^{\nu}_1 n^{\nu}_2..n^{\nu}_{N_{\nu}}),
\end{equation}
where $N_{\pi}$ and $N_{\nu}$ corresponds to the total number of single--particle orbitals for proton ($\pi$) and neutrons ($\nu$), respectively.
Now we can write the decomposed eigenstate as 
\begin{align}
    |\psi\rangle = 
    &\sum_{n_{\pi},n_{\nu}}
     c(n_{\pi},
       n_{\nu}) \nonumber |n_{\pi}\rangle
     \otimes |n_{\nu}\rangle.
\end{align}
Now the reduced density matrix can be calculated by tracing over one of the subspace indices:
\begin{equation}
    \rho^{(\pi)}=\sum_{n_{\nu}}\langle n_{\nu} | \langle n_{\pi} | \Psi \rangle \langle \Psi | n_{\pi}'  \rangle | n_{\nu} \rangle.
    \label{pn-RDM}
\end{equation}
Once the reduced density matrix is constructed, we can use the von Neumann entropy, which is a quantum extension of Shannon entropy, to measure the degree of correlation between the proton and neutron subspaces \cite{chuang}. The proton–neutron entanglement entropy is given by
\begin{equation}
S_{pn} = -\mathrm{Tr}[\rho^{(\pi)} \log_2 \rho^{(\pi)}],
\end{equation}
where a larger value of $S_{pn}$ indicates stronger entanglement between the proton and neutron parts of the nucleus, while $S_{pn}=0$ corresponds to a completely unentangled, separable state.

\subsection{Mutual information} \label{MI}
In addition to the equipartition entanglement, like the proton-neutron entanglement, a more refined understanding of quantum correlations can be achieved by studying orbital entanglement, i.e., correlations between individual single-particle orbitals or pairs of orbitals. Following the entanglement framework of Refs.\cite{Carl_2021,Boguslawski_2013} requires the construction of the reduced density matrices of one-orbital and two-orbitals (1-RDM and 2-RDM) for each orbital or pair of orbitals in the chosen single-particle basis, and calculating the corresponding entanglement entropy.

For the single-nucleon state $i$, we decompose the environment states $\mathbf{e}$ (composite state) as $\mathbf{e}_1 \in \{n_1,\ldots,n_{i-1}\}$ and 
$\mathbf{e}_2 \in \{n_{i+1},\ldots,n_N\}$, we can write the total wavefunction as 

\begin{equation}
    |\Psi\rangle \rightarrow |\Psi^{(n_i \mathbf{e})}\rangle 
    = \sum_{\mathbf{e}_1,n_i,\mathbf{e}_2} 
    c({\mathbf{e}_1,n_i,\mathbf{e}_2}) 
    | \mathbf{e}_1 \rangle \otimes |n_i\rangle \otimes |\mathbf{e}_2\rangle.
    \label{eq:wf_1_orb}
\end{equation}

The 1-RDM ${\rho^{(i)}}$ can now be calculated by tracing over the composite state,

\begin{equation}
    \rho^{(i)}
    = \mathrm{Tr}_{\mathbf{e}} 
    \left( 
    |\Psi^{(n_i\mathbf{e})}\rangle 
    \langle \Psi^{(n_i\mathbf{e})}|
    \right).
\end{equation}
This can be further simplified into a 2$\times$2 matrix consisting of only diagonal terms and written as

\begin{equation}
    \rho^{(i)} = \begin{pmatrix}
             1-\gamma_{ii} & 0 \\
             0 & \gamma_{ii}
            \end{pmatrix},
            \label{1-RDM}
\end{equation}
where $\gamma_{ii}=\langle \psi|\hat{a}^{\dagger}_i\hat{a}_i|\psi\rangle$ is occupation probability of the state $i$, the off-diagonal terms are zero because of particle-number conservation.

For the case of two-nucleon states $i$ and $j$, 
we split the environment states $\mathbf{e}$ as 
$\mathbf{e}_1 \in \{n_1,\ldots,n_{i-1}\}$, 
$\mathbf{e}_2 \in \{n_{i+1},\ldots,n_{j-1}\}$, and 
$\mathbf{e}_3 \in \{n_{j+1},\ldots,n_N\}$. 
The state is of the form
\begin{multline}
    |\Psi\rangle \rightarrow 
    |\Psi^{(n_i n_j \mathbf{e})}\rangle = \\ 
    \sum_{\mathbf{e}_1,n_i,\mathbf{e}_2,n_j,\mathbf{e}_3}
    c(\mathbf{e}_1,n_i,\mathbf{e}_2,n_j,\mathbf{e}_3)
    |\mathbf{e}_1\rangle \otimes |n_i\rangle \otimes 
    |\mathbf{e}_2\rangle \otimes |n_j\rangle \otimes |\mathbf{e}_3\rangle.
    \label{eq:wf_2_orb}
\end{multline}

The 2-RDM $\rho^{(ij)}$ can now be expressed as

\begin{equation}
    \rho^{(ij)} 
    = \mathrm{Tr}_{\mathbf{e}} 
    \left( 
    |\Psi^{(n_i n_j\mathbf{e})}\rangle 
    \langle \Psi^{(n_i n_j\mathbf{e})}|
    \right).
\end{equation}

The 2-RDM can now be written in matrix representation as 

\begin{multline}
\rho^{(ij)}= \\
    \begin{pmatrix}
1-\gamma_{ii}-\gamma_{jj}+\gamma_{ijij} & 0 & 0 & 0\\
0 & \gamma_{jj}-\gamma_{ijij}  & \gamma_{ji} & 0 \\
0 & \gamma_{ij} & \gamma_{ii}-\gamma_{ijij}  & 0 \\
0  & 0  & 0  & \gamma_{ijij}
\end{pmatrix},    
\end{multline}
where $\gamma_{ii}=\langle \psi|\hat{a}^{\dagger}_i\hat{a}_i|\psi\rangle$ and $\gamma_{jj}=\langle \psi|\hat{a}^{\dagger}_j\hat{a}_j|\psi\rangle$ denotes the occupation probability corresponding to $i$ and $j$ orbitals, $\gamma_{ijij}=\langle \psi|\hat{a}^{\dagger}_i\hat{a}^{\dagger}_j\hat{a}_i\hat{a}_j|\psi\rangle$ corresponds to the joint probability, and the non-diagonal term $\gamma_{ij}=\langle \psi|\hat{a}^{\dagger}_j\hat{a}_i|\psi\rangle$ is non-zero for like-nucleons due to species-number conservation. All other non-diagonal terms are zero because of particle-number conservation. The one-orbital and two-orbital entanglements become 

\begin{equation}
S_{i} = -\mathrm{Tr}\,[\rho_{i} \log_2 \rho_{i} ]
\end{equation}
and 

\begin{equation}
S_{ij} = -\mathrm{Tr}\,[\rho_{ij} \log_2 \rho_{ij} ],
\end{equation}
where all entropies and mutual information values are computed using logarithms to base 2.
Mutual information between orbitals $i$ and $j$ is defined as
\begin{equation}
I(i:j) = (S_i + S_j - S_{ij})(1 - \delta_{ij}),
\end{equation}
which quantifies the total correlations (classical + quantum) between the
two  \cite{chuang}. The factor of $(1 - \delta_{ij})$ is introduced to ensure the vanishing of the entanglement of a single-particle state with itself.

\subsection{Quantum Relative Entropy} \label{QRE}
In the previous sections, we utilized bipartition entanglement entropy and mutual information to quantify correlations within a single nuclear eigenstate. However, to study the structural evolution by quantifying the distinguishability between the ground states and excited states, a measure is required that can compare different quantum states. To achieve this, we introduce the frameworks of KLD and JSD, along with their quantum extensions.

The relative entropy, also known as the KLD, serves as a fundamental measure of distinguishability between two discrete probability distributions \cite{chuang}.
Let $P = \{p_k\}$ and $Q = \{q_k\}$ be two probability distributions representing the squared amplitudes of the Slater determinants in two different quantum states. The classical KLD is defined as
\begin{equation}
    D_{KL}(P \| Q) = \sum_{k} p_k \log_2 \left( \frac{p_k}{q_k} \right).
\end{equation}
However, it is asymmetric, $D_{KL}(P \| Q) \neq D_{KL}(Q \| P)$ and unbounded. To address these issues, we employ the classical JSD, a symmetrized and bound version of the KLD defined as
\begin{equation}
    D_{JS}(P \| Q) = \frac{1}{2} D_{KL}(P \| M) + \frac{1}{2} D_{KL}(Q \| M),
\end{equation}
where $M = \frac{1}{2}(P+Q)$ is the average distribution. This measure is bounded $0 \le D_{JS} \le 1$, and its square root constitutes a metric.

Now we will move towards the quantum extensions of KLD and JSD, which are appropriate mathematical formulations for the reduced density matrices corresponding to single-nucleon states given by Eqn.\ref{1-RDM} and proton-neutron partition given by Eqn.\ref{pn-RDM}. Let $\rho$ and $\sigma$ correspond to the reduced density matrices for the partitions being studied. The quantum generalization of the KLD, known as the quantum relative entropy, is given by
\begin{equation}
    D_{KL}(\rho \| \sigma) = \mathrm{Tr}[\rho (\log_2 \rho - \log_2 \sigma)].
\end{equation}
Analogous to the classical case, this measure is non-negative but not a true metric, nor is it bounded. So we will introduce the quantum extension of the JSD, which is given by 
\begin{equation}
    D_{JS}(\rho \| \sigma) = \frac{1}{2}D_{KL}(\rho\|\rho_M) + \frac{1}{2}D_{KL}(\sigma\|\rho_M),
\end{equation}
where $\rho_M = \frac{1}{2}(\rho + \sigma)$. In this work, the quantum JSD is applied to the proton–neutron reduced density matrices and the one-body reduced density matrices to quantify structural changes between nuclear states. A mode-resolved measure of distinguishability is obtained by evaluating the relative entropy for each single-nucleon orbital, and the total distinguishability is computed by summing over all orbital contributions, given by

\begin{equation}
    D^{tot}_{JS}(\rho \| \sigma) = \sum_{i} D^{i}_{JS}(\rho \| \sigma),
\end{equation}
where $i$ are the single-nucleon states.

This approach provides a physically interpretable measure of distinguishability between different nuclear eigenstates.

\section{RESULTS AND DISCUSSION} \label{results}

\subsection{Proton-neutron entanglement entropy}

In this subsection, we will discuss the proton-neutron entanglement entropy results in the $sdpf$ model space,
\begin{figure*}[!htbp]
    \centering

    \raisebox{0.32cm}{%
    \begin{subfigure}[t]{0.48\textwidth}
        \includegraphics[width=\textwidth]{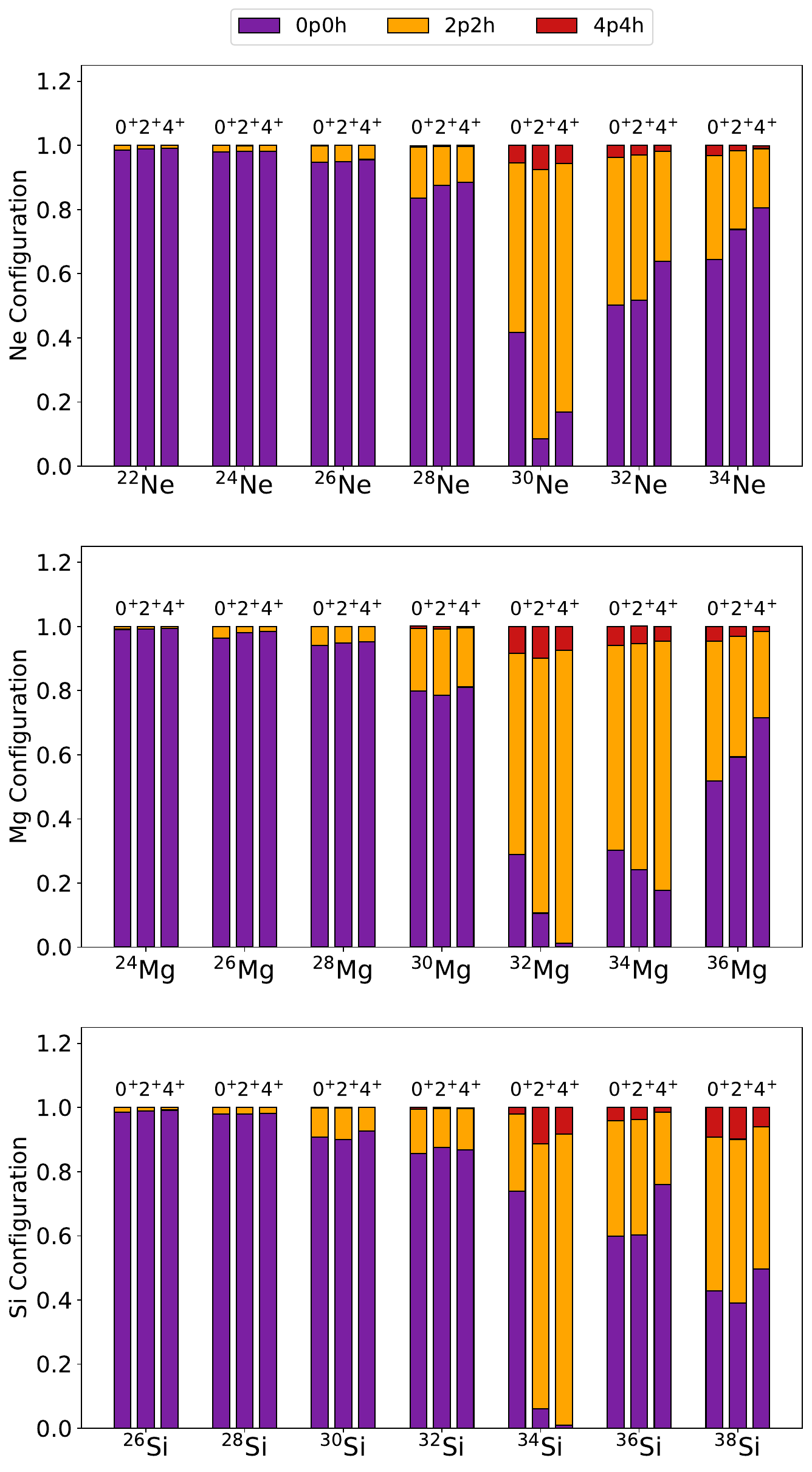}
        \label{fig:sub1}
    \end{subfigure}
    }
    \hfill
    \begin{subfigure}[t]{0.48\textwidth}
        \includegraphics[width=\textwidth]{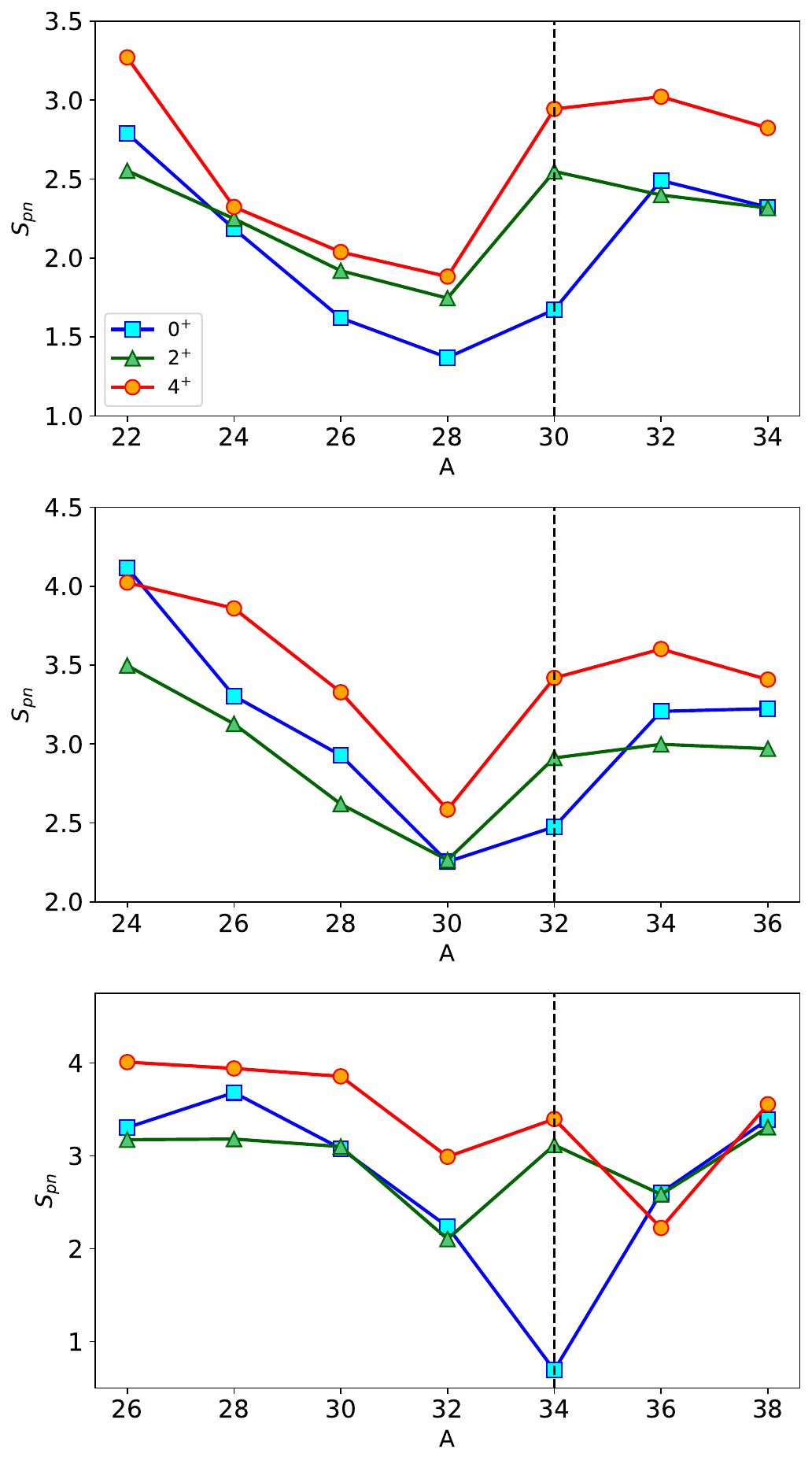}
        \label{fig:sub2}
    \end{subfigure}
    \caption{\small
Left panels: Configuration mixing for Ne, Mg, and Si isotopic chains, shown as the fractional contributions of normal (0p0h) and intruder (2p2h and 4p4h) configurations in the $sdpf$ model space for the lowest $0^+$, $2^+$, and $4^+$ states. 
Right panels: Corresponding proton-neutron entanglement entropy $S_{pn}$ for the same nuclei and states as a function of mass number. The vertical dashed lines indicate the position of the $N=20$ shell closure.}

    \label{fig:pn_ent_Ne_Mg_Si}
\end{figure*}
specifically in the island of inversion region. This phenomenon arises from the weakening of the $N=20$ shell gap and the accompanying increase in correlation energy, which allows intruder configurations to dominate over normal ones.  
  As shown in Fig.~\ref{fig:pn_ent_Ne_Mg_Si}, the $0^{+}$ states of Ne
and Mg isotopes are primarily composed of normal configurations for $N \leq 18$.  However, beginning with ${}^{30}$Ne and ${}^{32}$Mg, 2p2h intruder configurations become dominant, marking the onset of the IoI.

We have computed and plotted the proton-neutron entanglement entropy for the isotopic chains of Ne (${}^{22-34}$Ne), Mg (${}^{24-36}$Mg), Si (${}^{26-38}$Si), for the lowest $0^+$, $2^+$, and $4^+$ states. As previously shown in Ref.~\cite{CJ_2023}, the entropy generally reaches its maximum at $N=Z$ and decreases with increasing neutron number when only a single major shell is considered, reflecting the reduced neutron configurations. However, in the present study, when the $sdpf$ model space is considered, cross-shell excitations across the $N=20$ shell gap enhance configuration mixing, leading to an increase in entropy. 

From Fig.\ref{fig:pn_ent_Ne_Mg_Si}, we observe that the entropy decreases steadily toward neutron-rich nuclei in all three isotopic chains. However, at $N=20$, the entropy rises again in the Ne and Mg chains, indicating an increase in correlation between the proton-neutron partitions of the wavefunction. This rise can be attributed to neutron excitations into the $pf$ shell, where the occupancy of the $d_{3/2}$ orbital 
saturates, and the increasing occupancy of the $pf$ orbitals leads to a broader distribution of neutrons and, consequently, higher entropy.
\begin{figure}[t]
    \centering
    \includegraphics[width=0.5\textwidth]{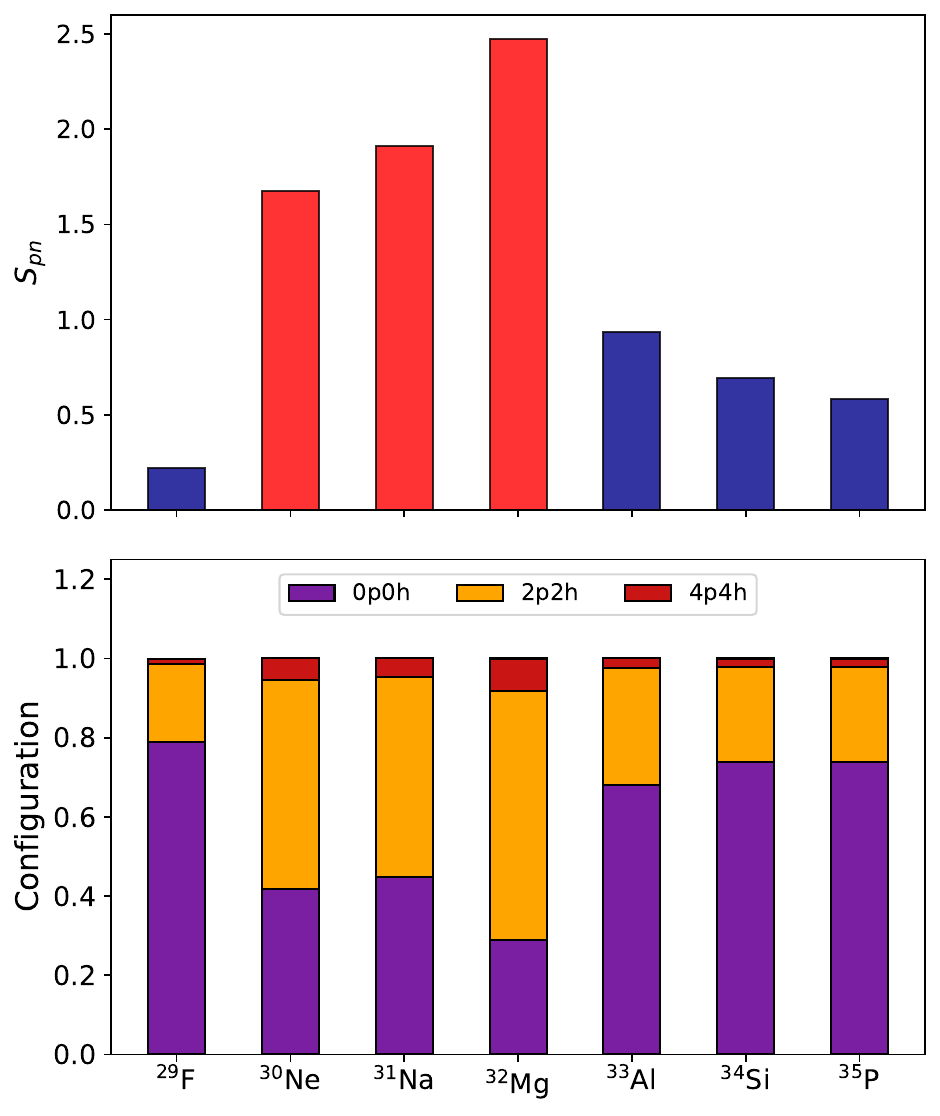}
    \caption{\small
Proton-neutron entanglement entropy $S_{pn}$ (upper panel) and corresponding configuration (lower panel) for the $N=20$ isotones from $^{29}$F to $^{35}$ P. The red bars correspond to the nuclei under the IoI. The configuration mixing is shown as the fractional contributions of normal (0p0h) and intruder (2p2h and 4p4h) configurations.}

    \label{fig:single_diagram}
\end{figure}

For the Si isotopic chain, the ground state entropy plummets to a very low value, indicating weak proton-neutron correlations. This behavior can be explained by ${}^{34}$Si lying outside the IoI, as well as a subshell closure of the $d_{5/2}$ orbital in the proton valence space. For the $2^+$ and $4^+$ states, the 2p2h configuration dominates, resulting in an increase in entropy. The increase observed for ${}^{36}$Si and ${}^{38}$Si is consistent with the addition of valence neutrons in the $pf$ shell, which enhances the proton-neutron correlations.

In Fig.\ref{fig:single_diagram}, we present a direct comparison between the proton-neutron entanglement entropy and the configuration mixing for the $N=20$ isotones. An increase in intruder configuration, which is dominated by 2p2h in ${}^{30}$Ne, ${}^{31}$Na, ${}^{32}$Mg, corresponds to an increase in entropy, and lowers as normal configuration becomes a dominant configuration again. 

{\color{black}We note that while previous studies \cite{CJ_2023} within a single major shell reported a reduction of proton--neutron entanglement entropy toward neutron-rich nuclei, the enhancement observed here around $N=20$ is associated with the island-of-inversion region, where cross-shell excitations and intruder configurations strengthen proton--neutron correlations.} This analysis demonstrates that proton-neutron entanglement entropy serves as a probe for studying correlations, such as quadrupole correlations, which are key drivers of the emergence of the IoI.

\vspace{-0.2cm}
\begin{figure}[h]
    \centering
    \includegraphics[width=0.52\textwidth]{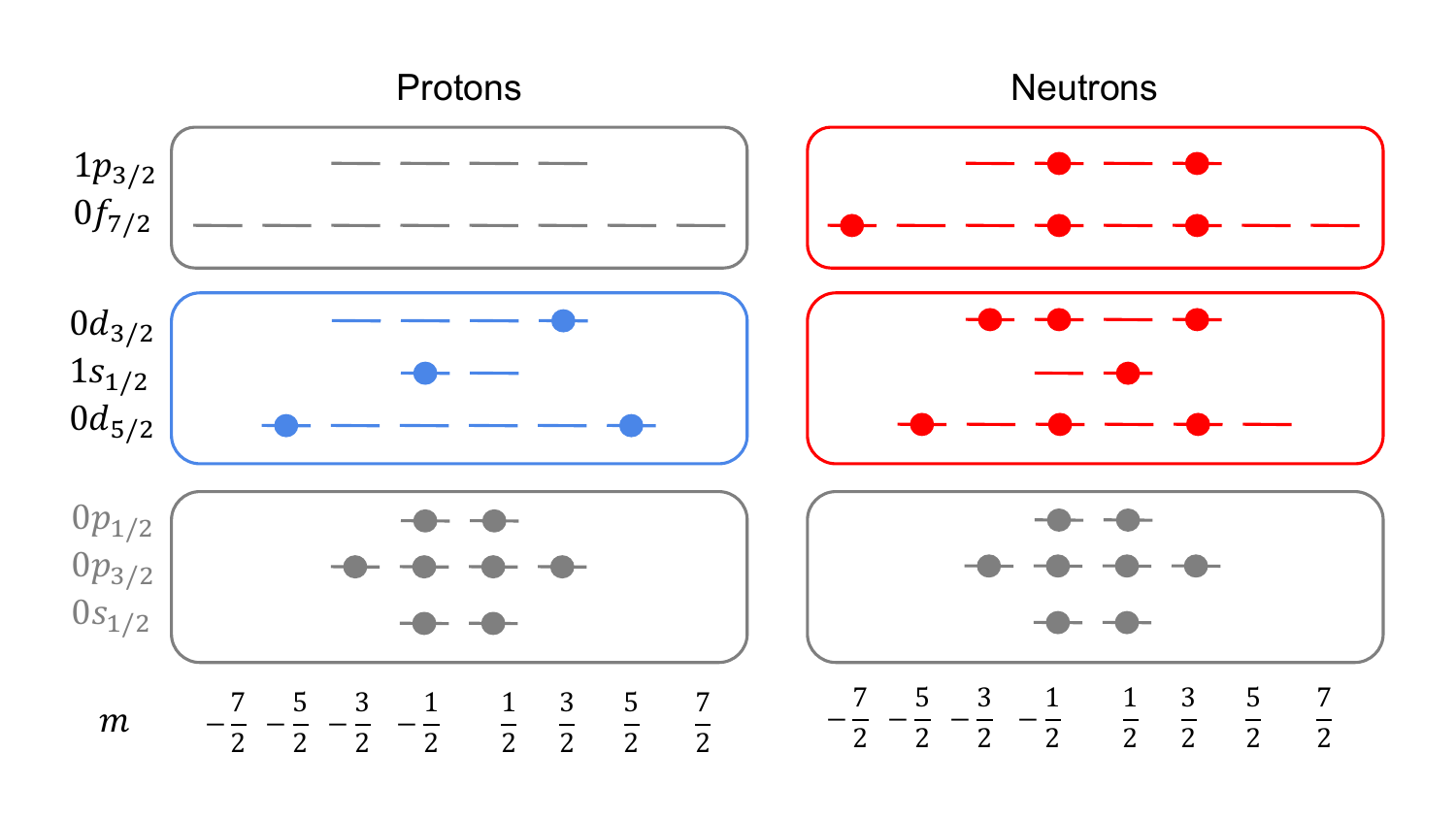}
    \caption{\small
    We have illustrated the valence spaces considered for our calculations, consisting of the $sd$ shell for protons (blue) and the $sd$ and $pf$ shells for neutrons (red). The fully filled $^{16}$O core and the restricted $pf$ shell for protons are represented in grey. We have demonstrated one of the configurations of $^{32}$Mg.}
    \label{fig:sdpf_model}
\end{figure}

\subsection{Mutual information} 
\begin{figure*}[t]
    \centering
    \includegraphics[width=0.82\textwidth]{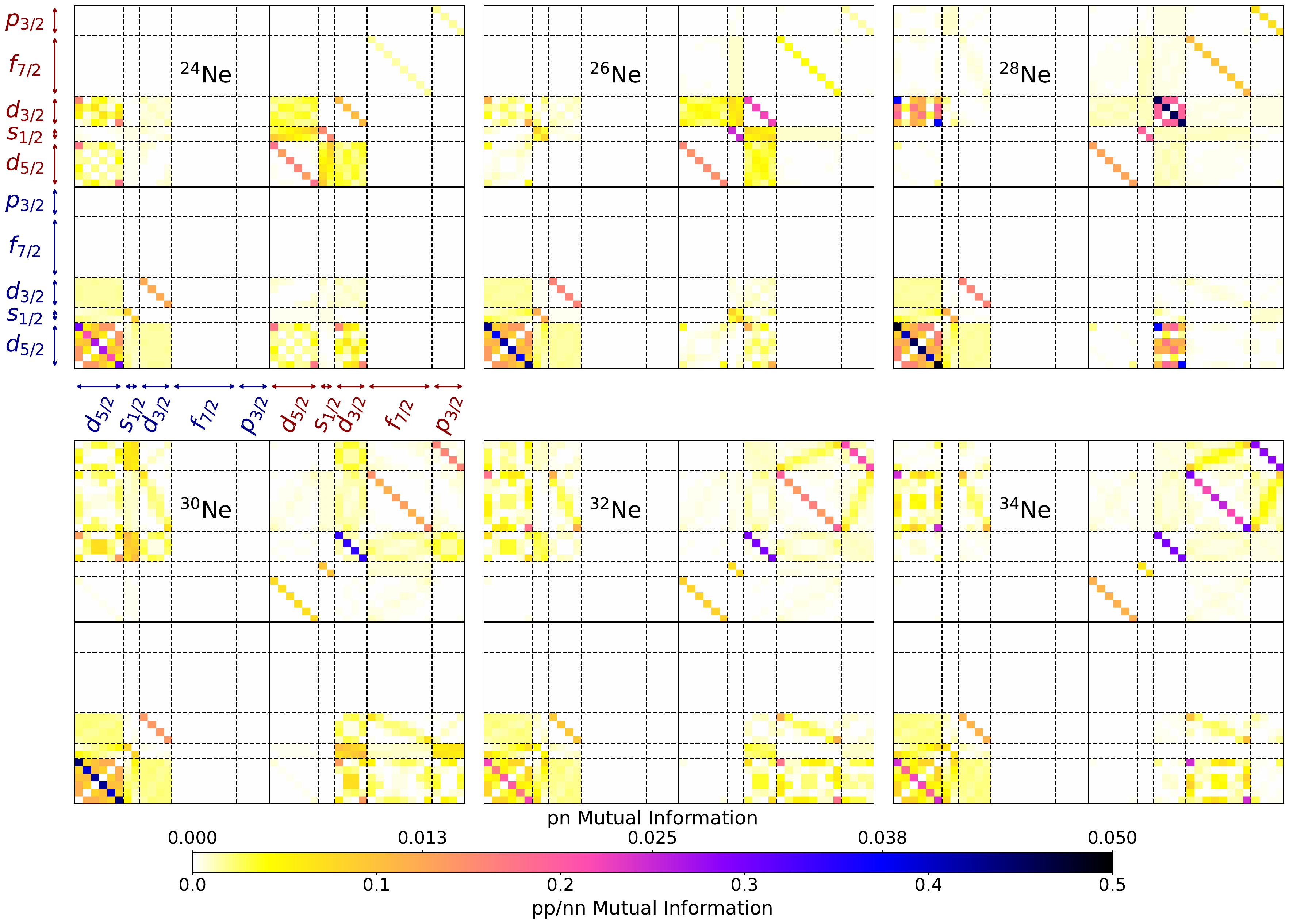}
    \caption{\small
Mutual information for the first $0^+$ states of $^{24-34}$Ne in the $sdpf$ model space. Red (blue) arrows denote neutron (proton) orbitals, with magnetic substates ordered from $-j$ to $+j$. {\color{black}Solid lines separate the proton--proton (bottom--left), neutron--neutron (top--right), proton-neutron (top--left), and neutron--proton (bottom--right), while dashed lines indicate the boundaries between subshells.}}
    \label{fig:Ne_MI_0+}
\end{figure*}

\begin{figure*}[htbp]
    \centering
    \includegraphics[width=0.82\textwidth]{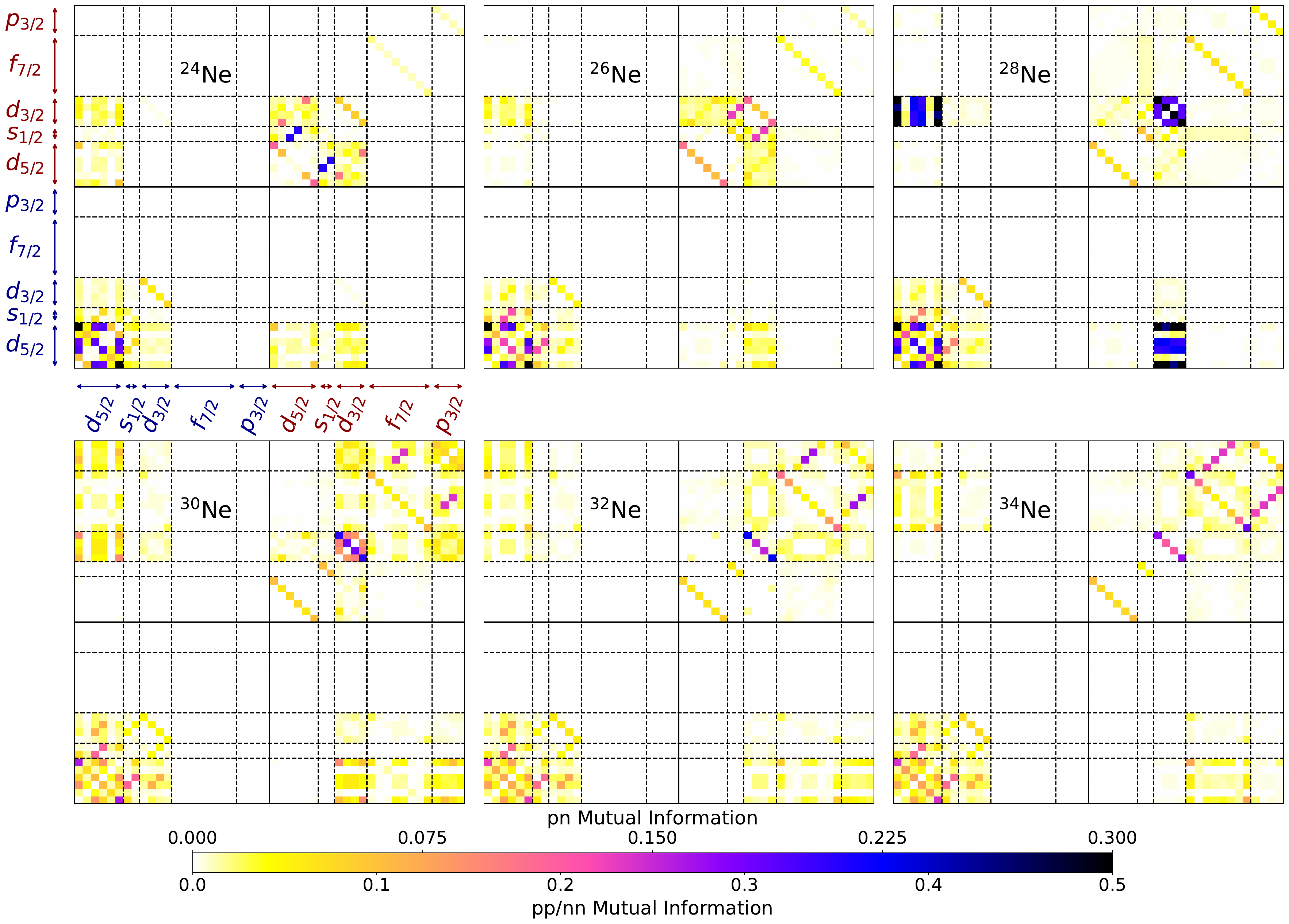}
    \caption{Mutual information for the first $2^+$ states of $^{24-34}$Ne in the $sdpf$ model space. Same scheme as followed in Fig.\ref{fig:Ne_MI_0+}. The magnitude of the colorbar for the pn sectors differs from that of the $0^+$ states.}
    \label{fig:Ne_MI_2+}
\end{figure*}

\begin{figure*}[htbp]
    \centering
    \includegraphics[width=0.82\textwidth]{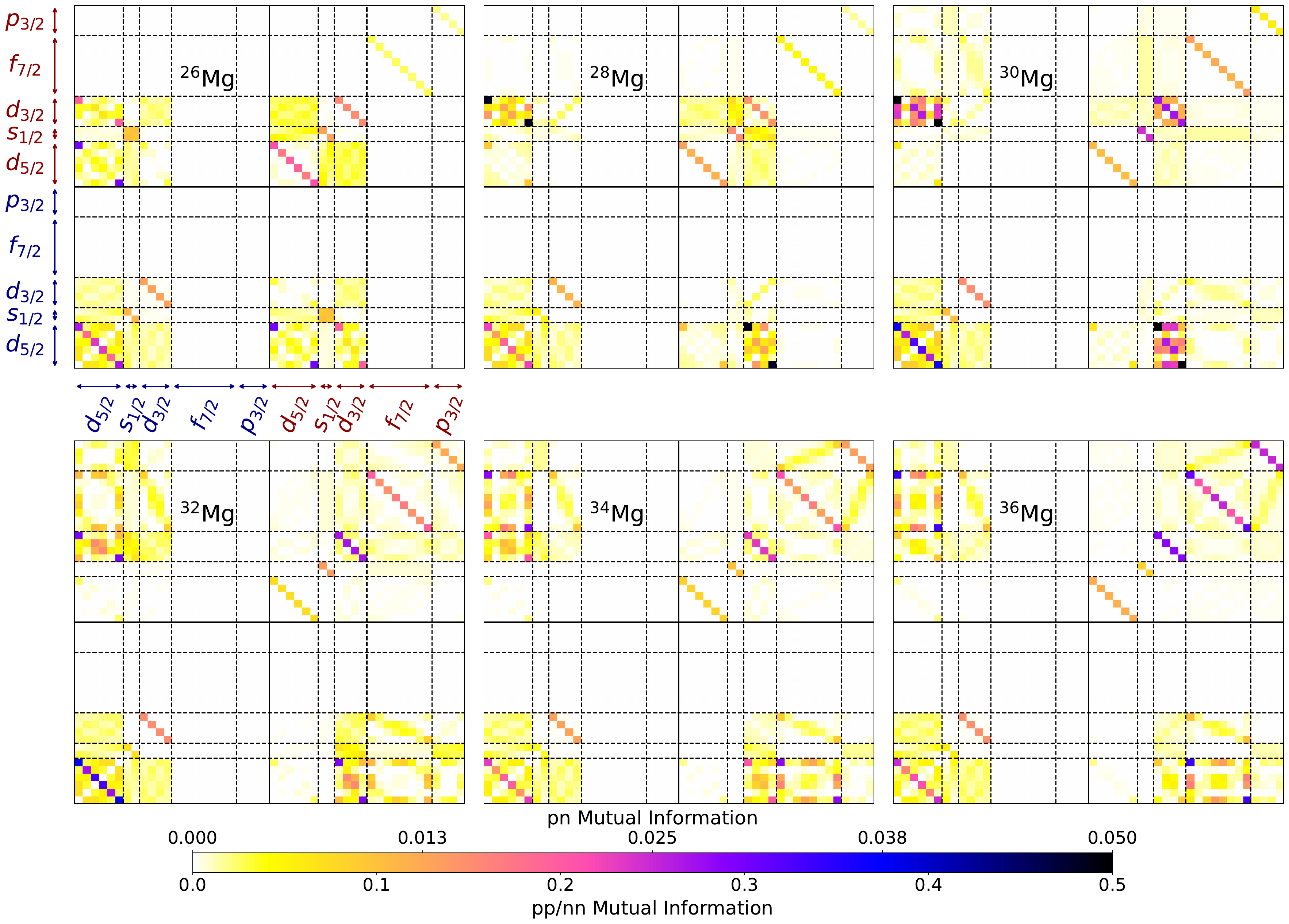}
    \caption{Same as Fig.\ref{fig:Ne_MI_0+}, but for $^{26-36}$Mg isotopic chain.}
    \label{fig:Mg_MI_0+}
\end{figure*}

\begin{figure*}[htbp]
    \centering
    \includegraphics[width=0.82\textwidth]{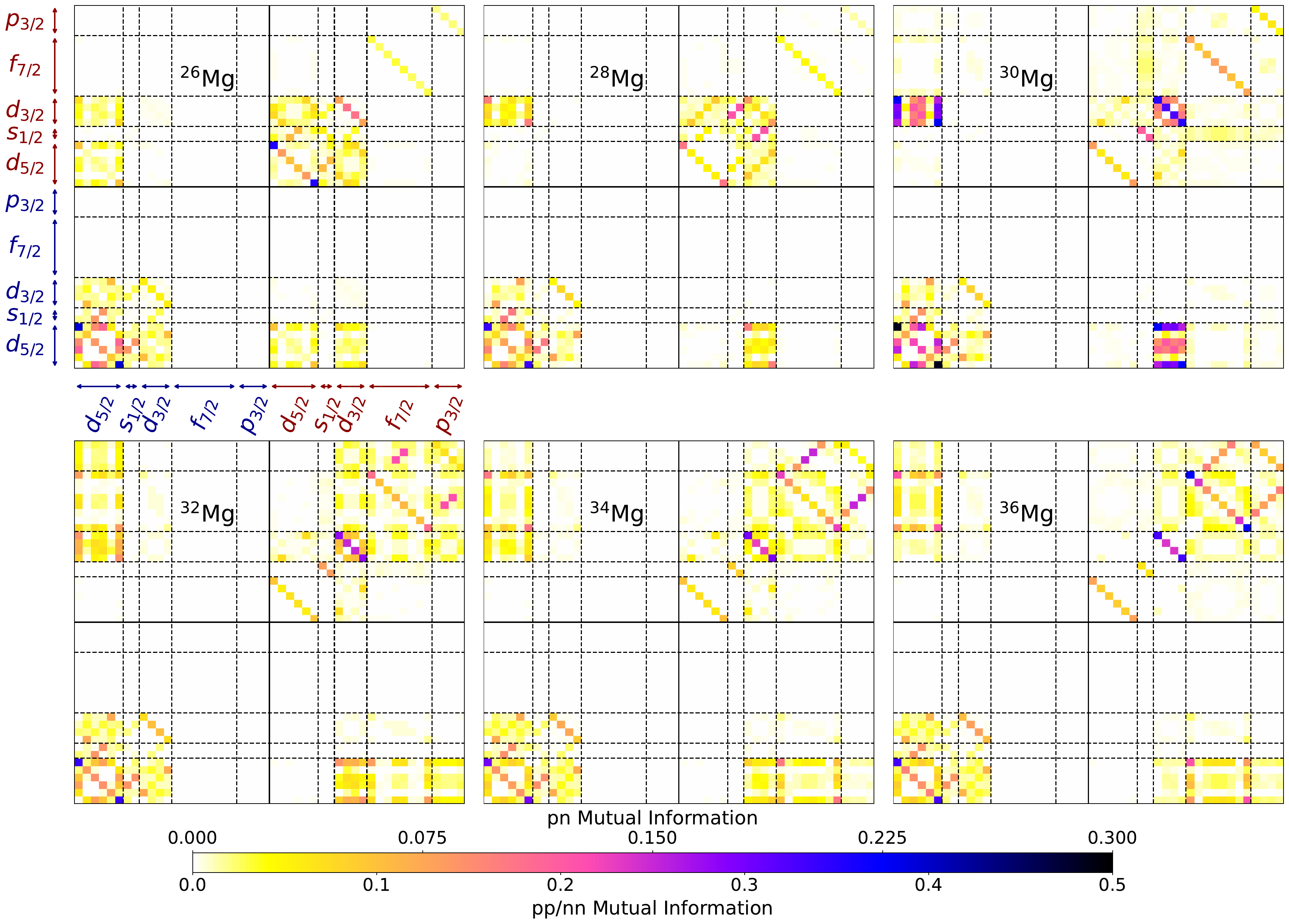}
    \caption{Same as Fig.\ref{fig:Ne_MI_2+}, but for $^{26-36}$Mg isotopic chain.}
    \label{fig:Mg_MI_2+}
\end{figure*}

\begin{figure*}[htbp]
    \centering
    \includegraphics[width=0.82\textwidth]{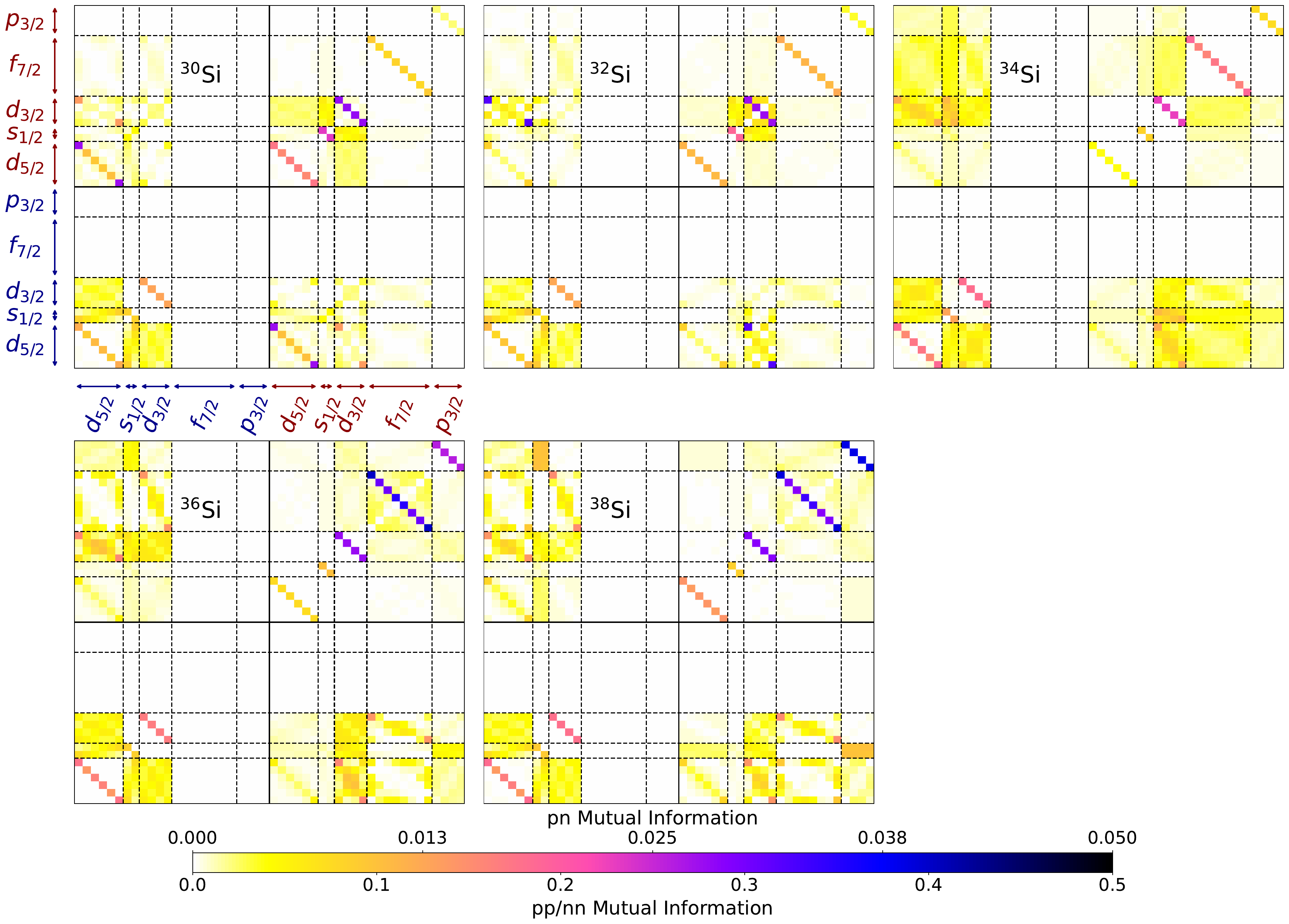}
    \caption{Same as Fig.\ref{fig:Ne_MI_0+}, but for $^{30-38}$Si isotopic chain.}
    \label{fig:Si_MI_0+}
\end{figure*}

\begin{figure*}[htbp]
    \centering
    \includegraphics[width=0.82\textwidth]{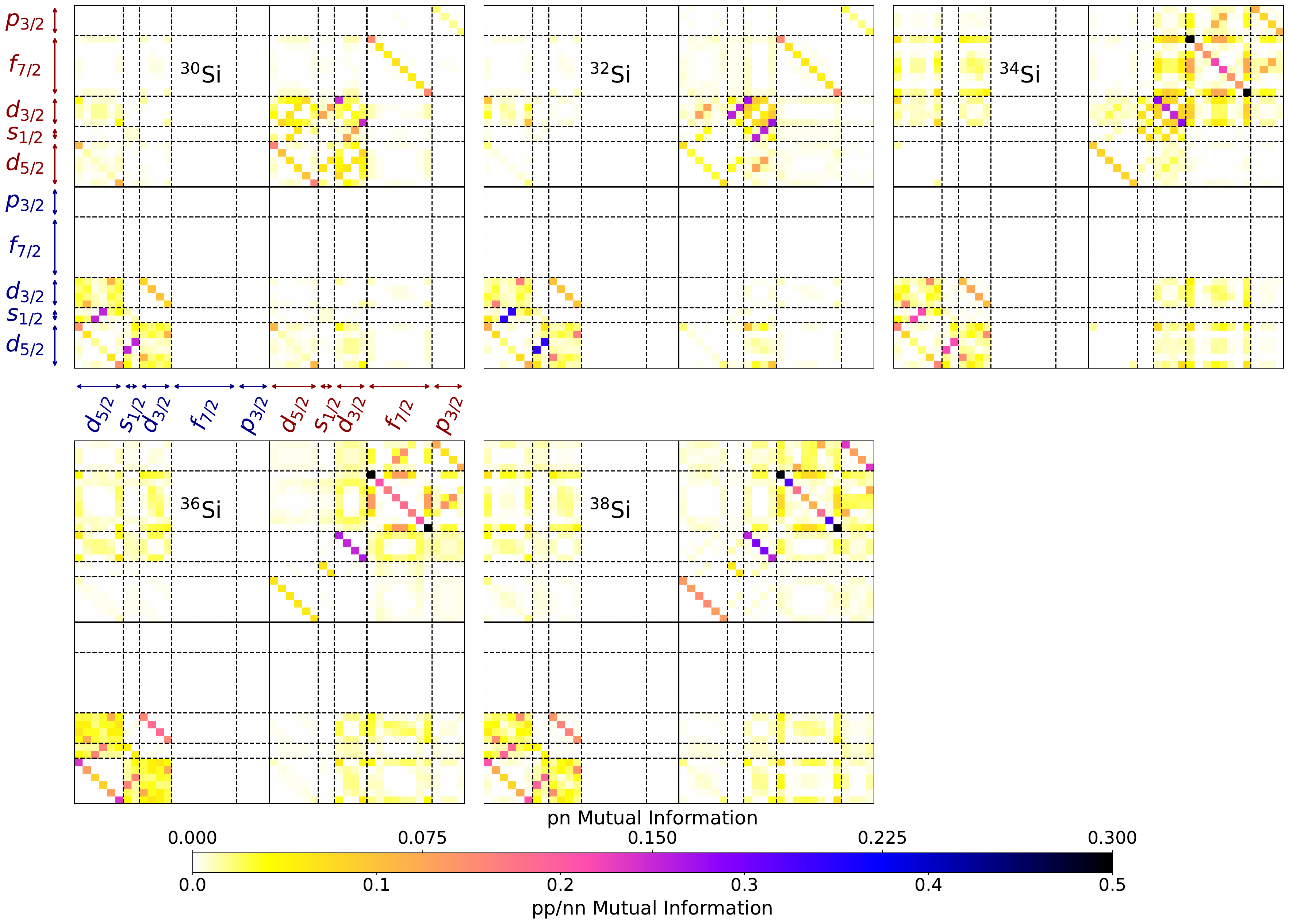}
    \caption{Same as Fig.\ref{fig:Ne_MI_2+}, but for $^{30-38}$Si isotopic chain.}
    \label{fig:Si_MI_2+}
\end{figure*}

 The valence space consists of 12 proton single-particle states and 24 neutron single-particle states in the M-scheme many-body basis, which is presented in Fig. \ref{fig:sdpf_model}. 
We restrict protons to the $sd$ model space and neutrons to the $sdpf$ model space. 
The mutual information results, which include both quantum and classical correlations between the pairs of single-particle states ($i,j$), are shown in Figs. \ref{fig:Ne_MI_0+}-\ref{fig:Si_MI_2+}. The solid black lines divide the mutual information plot into four sections: the bottom-left section corresponds to proton–proton correlations, and the top-left and bottom-right sections represent proton–neutron correlations,
 and the top-left and bottom-right
sections represent proton–neutron and neutron-proton correlations, respectively.
 The dashed lines separate the subshells in the order $d_{5/2}$, $s_{1/2}$, $d_{3/2}$, $f_{7/2}$, and $p_{3/2}$. Each subshell is constructed using the third component of the angular momentum, consisting of $2j+1$ orbitals, where $j$ is the angular momentum of that subshell. 
To better interpret the results, we classify the correlations according to the orbitals involved. Isovector correlations correspond to pairs of orbitals with opposite magnetic projections $m$ ($m_1 + m_2 = 0$) within the same subshell. {\color{black} These correlations are dominated by the isovector pairing part of the interaction, hence the name.} Non-isovector correlations correspond to off-diagonal correlations within the same subshell, and finally, cross-orbital correlations arise between orbitals belonging to different subshells. To clearly visualize the evolution of proton–neutron correlations in the ground states, the scale for the proton–neutron sector has been reduced by an order of magnitude compared to the like-particle sectors. However, the scale of the proton-neutron sector for the $2^+$ state of each isotopic chain is restricted to $0.3$, which is comparable to the like-particle scale of $0.5$, suggesting a large proton-neutron correlation compared to the ground states.

We begin our analysis with Fig. \ref{fig:Ne_MI_0+}, where the proton-proton sector of ${}^{24}$Ne shows a high mutual information between the $d_{5/2}$ orbitals, especially between orbitals with opposite angular-momentum projection $m$, a feature that will be very common in the ground states of the even-even isotopes. The neutron-neutron sector also displays strong isovector correlations for $d_{5/2}$, $s_{1/2}$, and $d_{3/2}$ orbitals, along with moderate cross-orbital correlations. Very weak correlations are observed in the proton-neutron sector, which are approximately an order of magnitude smaller than those in the like-particle sectors. Similar structures are observed in ${}^{26}$Ne, with stronger correlations between opposite $m$ orbitals. In ${}^{28}$Ne, the correlations among the $d_{3/2}$ orbitals of the neutron-neutron sector become more pronounced, as well as proton-neutron correlations between proton $d_{5/2}$ and neutron $d_{3/2}$ orbitals grow stronger. This can be attributed to the $N=20$ shell gap, which restricts the excitation of neutrons in the $pf$ shell, thereby enhancing the correlations in $d_{3/2}$ orbitals. A common feature observed in ${}^{30}$Ne, ${}^{32}$Ne, and ${}^{34}$Ne is an increase in the proton-neutron correlations, driven by the collapse of $N=20$ shell gap which causes an increase in mutual information between proton $d_{5/2}$ orbital and neutron $f_{7/2}$ and $p_{3/2}$ orbitals, the correlations are still weak by an order of magnitude as compared to proton-proton and neutron-neutron. We can also observe traces of correlations between neutron $d_{3/2}$ orbitals and $pf$ shell orbitals in ${}^{30}$Ne. Now we will analyze the $2^+$ states for the Ne isotopic chain as shown in Fig. \ref{fig:Ne_MI_2+}. Some of the noticeable patterns in the excited states are the weakening of isovector correlations in the proton-proton sectors and equally strong non-isovector correlations. Also, the proton-proton correlations decrease progressively towards neutron-rich nuclei. In the neutron-neutron sector, the isovector pairing is stronger or on par with the non-isovector correlations for nuclei, except for ${}^{24}$Ne. The proton-neutron correlations for ${}^{24}$Ne, ${}^{26}$Ne, and ${}^{28}$Ne nuclei are restricted between the proton $d_{5/2}$ orbitals and neutron $d_{5/2}$, $s_{1/2}$, and $d_{3/2}$ orbitals. However, for the nuclei within the IoI, the mutual information shifts to $d_{3/2}$, $f_{7/2}$, and $p_{3/2}$ due to cross-shell excitation. 

A similar analysis has been performed for the Mg isotopic chain, as shown in Fig. \ref{fig:Mg_MI_0+} and Fig. \ref{fig:Mg_MI_2+}. The ground states of the Mg isotopes (Fig. \ref{fig:Mg_MI_0+}) exhibit similar features to those of the Ne isotopes, where isovector correlations dominate for both the proton–proton and neutron–neutron sectors. This can be explained by interactions that favor isovector pairings. The proton-proton correlations in the $d_{5/2}$ orbitals are slightly weaker in the Mg chain compared to the Ne chain. The proton-neutron correlations also exhibit similar characteristics, with the mutual information being restricted to the $sd$ valence shell until ${}^{30}$Mg. Beyond which the correlations increase between proton orbitals and neutron $d_{3/2}$, $f_{7/2}$, and $p_{3/2}$ orbitals, indicating the neutron excitations and increase in proton-neutron correlation, which are still weaker by an order of magnitude when compared to like-particle correlations. In the excited $2^+$ states (Fig. \ref{fig:Mg_MI_2+}), the strength of non-isovector correlations is comparable to that of the isovector correlations in the proton-proton sector. In the neutron-neutron sectors as well, the non-isovector correlations are prominent, especially between the $f_{7/2}$ and $p_{3/2}$ orbitals for ${}^{32}$Mg, ${}^{34}$Mg, and ${}^{36}$Mg.   

We now turn to the Si isotopic chain in Fig. \ref{fig:Si_MI_0+} and Fig. \ref{fig:Si_MI_2+}. In the ground states (Fig. \ref{fig:Si_MI_0+}) of this isotopic chain, the proton-proton sector exhibits a strong isovector correlation. However, the like-orbital correlations for the proton-proton and neutron-neutron sectors exhibit negligible non-isovector correlations, as observed in the Ne and Mg isotopic chains. Moderate correlations are observed between proton $d_{5/2}$, $s_{1/2}$, and $d_{3/2}$ orbitals for the entire isotopic chain.  In the neutron-neutron sector, the correlations are dominated by isovector pairing correlations for every isotope, where the correlations in the $pf$ shell increase as the nuclei become more neutron-rich. The proton-neutron correlations are weak, and an opposite $m$ correlation structure in the ${}^{30}$Si is observed between proton and neutron $d_{5/2}$ orbitals. Some moderate cross-shell correlations are observed between proton $d_{5/2}$ orbitals and neutron $d_{3/2}$ orbitals in $^{34-38}$Si nuclei. For the excited states of the entire isotopic chain (Fig. \ref{fig:Si_MI_2+}), the neutron-neutron sector has non-isovector correlations comparable to those of isovector pairs. The proton-proton sector has isovector pairing correlations comparable to cross-shell correlations between $d_{5/2}$, $s_{1/2}$, and $d_{3/2}$ orbitals. In the proton-neutron sector, the correlations are weaker than those of Ne and Mg proton-neutron correlations.

\begin{figure*}[t]
    \centering

    \begin{subfigure}[b]{0.49\textwidth}
        \includegraphics[width=\textwidth]{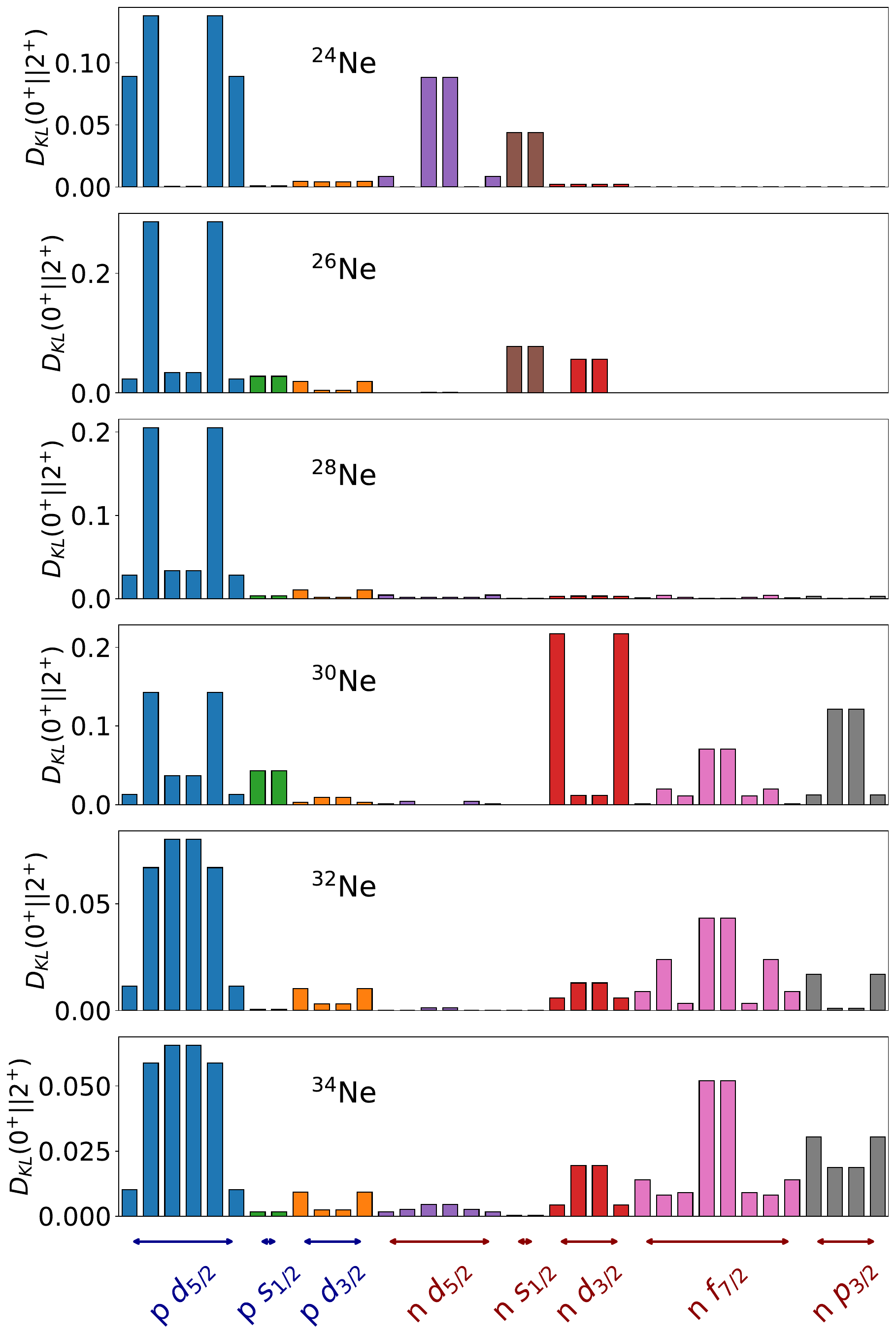}
        \label{fig:sub1}
    \end{subfigure}
    \hfill
    \begin{subfigure}[b]{0.49\textwidth}
        \includegraphics[width=\textwidth]{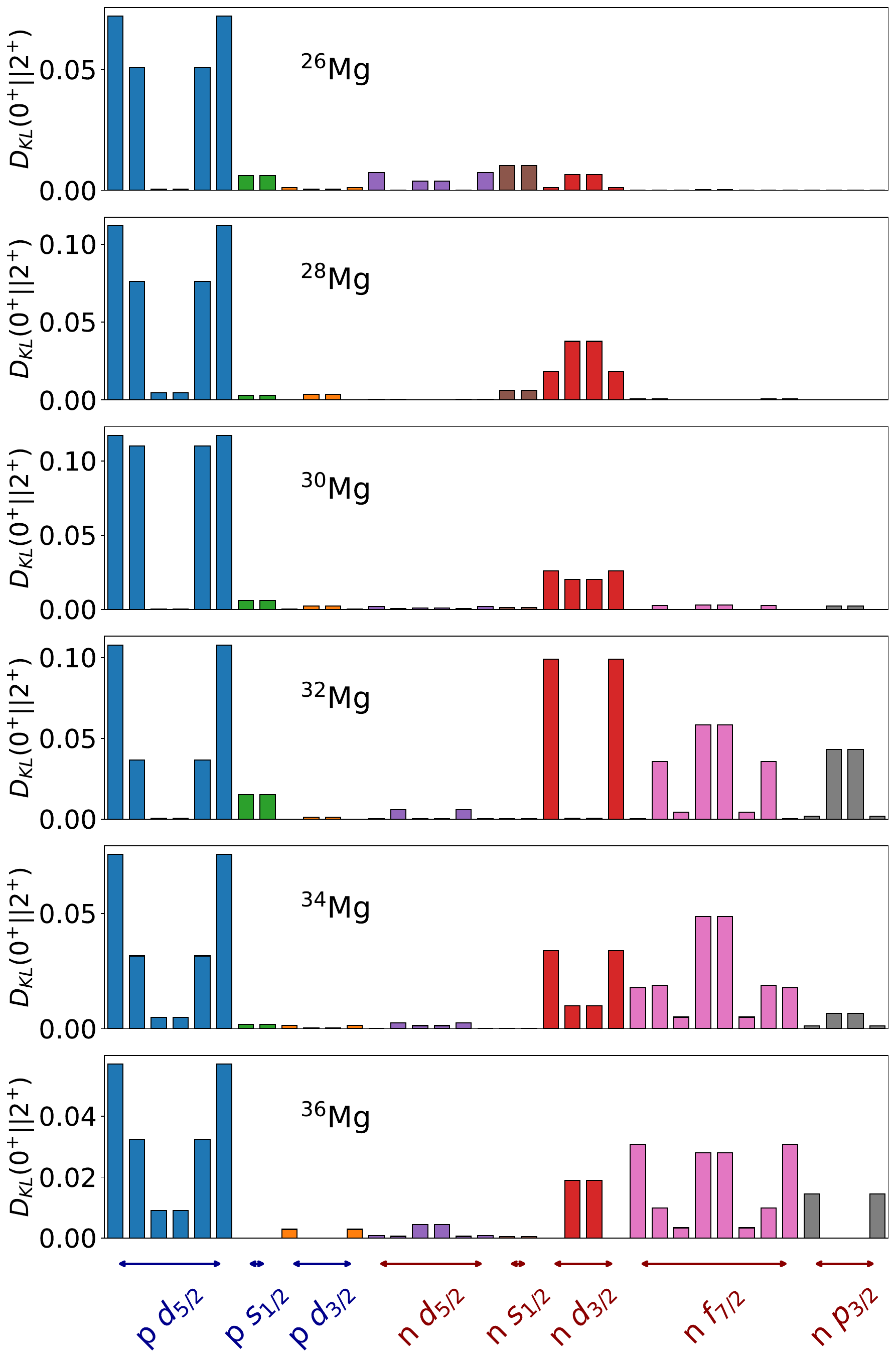}
        \label{fig:sub2}
    \end{subfigure}
    \caption{\small
    Left Panels: Mode-resolved quantum relative entropy, quantified using the Kullback--Leibler divergence $D_{KL}$, between the ground state $0^+$ and first $2^+$ excited state for $^{24-34}$Ne isotopic chain. Blue (red) arrows denote proton (neutron) single-particle orbitals, with magnetic substates ordered from $-j$ to $+j$. 
    Right Panels:  Mode-resolved quantum relative entropy between the ground state $0^+$ and first $2^+$ excited state for $^{26-36}$Mg isotopic chain.}

    \label{fig:Mode_KLD_Ne_Mg}
\end{figure*}

\begin{figure*}[t]
    \centering

    \begin{subfigure}[b]{0.49\textwidth}
        \includegraphics[width=\textwidth]{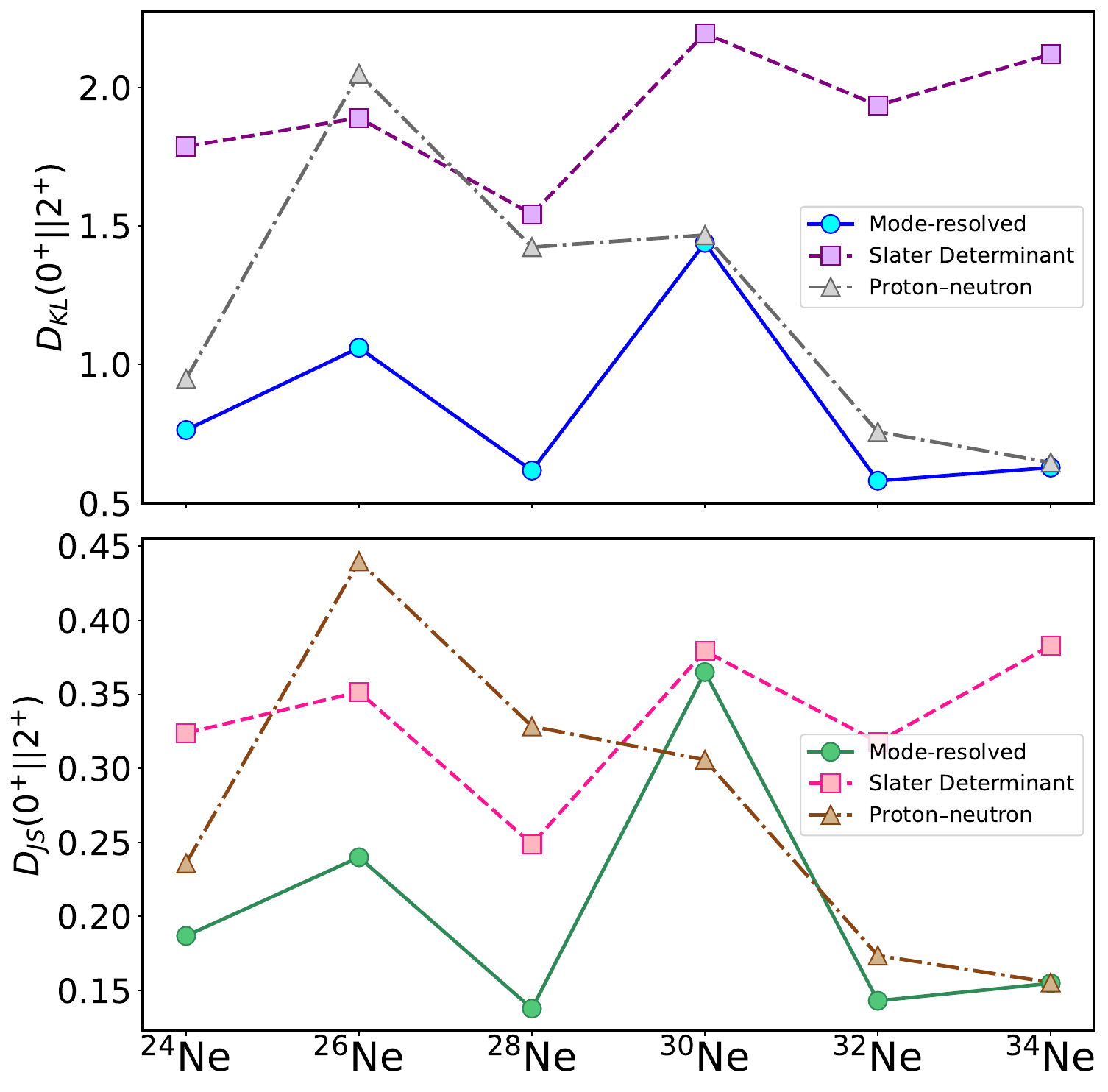}
        \label{fig:sub1}
    \end{subfigure}
    \hfill
    \begin{subfigure}[b]{0.49\textwidth}
        \includegraphics[width=\textwidth]{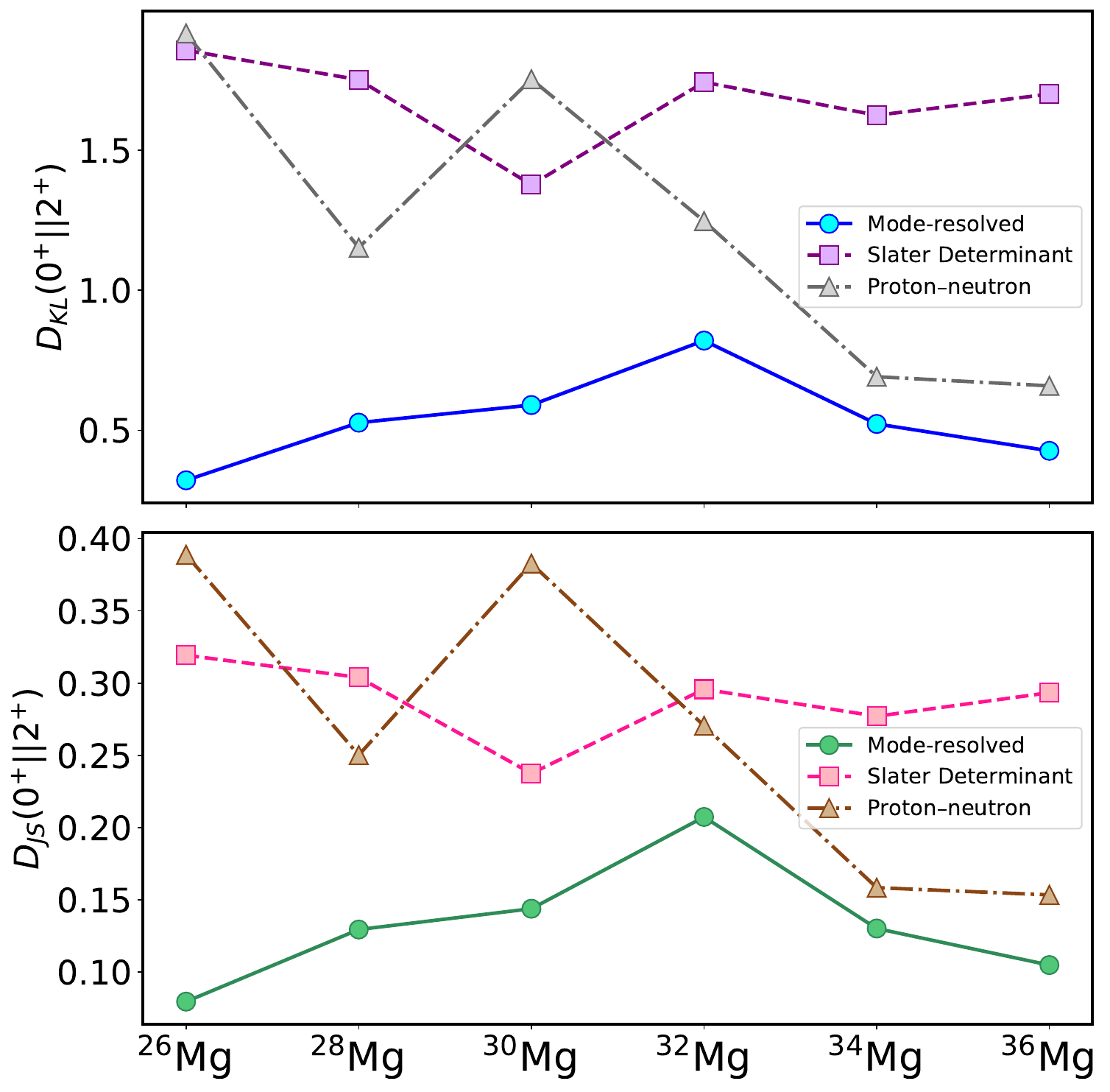}
        \label{fig:sub2}
    \end{subfigure}
    \caption{\small
    Left Panels: Kullback-Leibler divergence $D_{KL}$ and Jensen-Shannon divergence $D_{JS}$ between the ground states ($0^+$) and the first $2^+$ states for the $^{24-34}$Ne isotopic chain for the mode-resolved, Slater determinant and proton-neutron partitions.
    Right Panels: Same as the Left Panels, but for the $^{26-36}$Mg isotopic chain.}

    \label{fig:KLD_JSD_Ne_Mg}
\end{figure*}

\subsection{Quantum relative entropy}

The proton–neutron entanglement entropy and mutual information discussed in the previous sections quantify correlations arising from partitions within a single nucleus. To extend this analysis and compare the quantum informational properties of two distinct {\color{black} quantum states}, quantum relative entropy serves as a powerful tool. It provides a quantitative measure of distinguishability between quantum states. In this section, we present quantum relative entropy results for three distinct types of partitions. The first is mode-resolved, where we compare individual single-particle states (often referred to as “modes” in quantum information theory). The second compares the Slater determinant components of the two nuclear states under consideration. The third examines the proton–neutron partition of the model space. For the mode-resolved relative entropy, we also show the individual contributions from each single-particle state. For the Ne, Mg, and Si isotopes, the quantum relative entropy is evaluated between the $0^+$ ground state and the first $2^+$ excited state. For the $N=20$ isotones, the calculations are carried out between their respective ground and first excited states for odd-even nuclei, and $0^+$ and $2^+$ for even-even nuclei.

We begin by discussing the quantum relative entropy between single-particle states for the Ne isotopic chain, as shown in Fig.~\ref{fig:Mode_KLD_Ne_Mg}. In these calculations, we have used the $sd$ model space for protons and the $sd$ model space, together with the $f_{7/2}$ and $p_{3/2}$ orbitals, for neutrons. For the proton orbitals, the maximum contribution comes from the $d_{5/2}$ orbitals, showing that they are the most distinguishable orbitals in the proton model space. The quantum relative entropy evolves for the Ne chain as we move towards neutron-rich nuclei. Now we will analyze the neutron orbitals, for ${}^{24}$Ne, both the $d_{5/2}$ and $s_{1/2}$ orbitals contribute, while in ${}^{26}$Ne the $d_{3/2}$ and $s_{1/2}$ neutron orbitals take part, their contribution are not significant compared to the proton $d_{5/2}$ orbitals. The neutron orbitals for ${}^{28}$Ne have negligible contributions to the quantum relative entropy. This can be explained by the $N=20$ shell gap, which confines the configurations within the $sd$ shell, making the $0^+$ and $2^+$ states almost indistinguishable. For the heavier isotopes ${}^{30}$Ne, ${}^{32}$Ne, and ${}^{34}$Ne, most of the contribution to the quantum relative entropy comes from the neutron $d_{3/2}$, $f_{7/2}$, and $p_{3/2}$ orbitals, it is also worth noting that the magnitude of these contributions for ${}^{32}$Ne and ${}^{34}$Ne are low compared to $^{30}$Ne. The leftmost panels in Fig. \ref{fig:KLD_JSD_Ne_Mg} correspond to the quantum relative entropy of the Ne isotopic chain, where the top panel is KLD and the bottom panel is JSD. The defining feature of the mode-resolved relative entropy in both cases is the decrease in entropy in ${}^{28}$Ne, which results from very small contributions from the neutron orbitals, meaning the $0^+$ and $2^+$ states are nearly indistinguishable. This is followed by the highest relative entropy of ${}^{30}$Ne, which can be explained by the higher percentage of intruder configurations in the $2^+$ state as compared to the $0^+$ state, causing $d_{3/2}$, $f_{7/2}$, and $p_{3/2}$ orbitals to contribute. The last two nuclei of the {\color{black}isotopic chain} have low entropy, which can be explained by the same percentage of intruder configurations for $0^+$ and $2^+$ states for the corresponding nuclei. The relative entropy corresponding to the Slater determinant shows deviation in patterns in KLD and JSD for ${}^{24}$Ne and ${}^{26}$Ne, which can be attributed to KLD not being symmetric and not a true distance measure. But it does match the pattern observed in JSD from ${}^{28}$Ne, which falls, and rises for ${}^{30}$Ne. The relative entropy calculated for the proton–neutron partition also follows a similar pattern in both measures. It starts low for {\color{black}${}^{24}$Ne}, peaks around ${}^{26}$Ne, and then gradually decreases as we move toward the neutron-rich side of the isotopic chain.

\begin{figure*}[t]
    \centering

    \begin{subfigure}[b]{0.49\textwidth}
        \includegraphics[width=\textwidth]{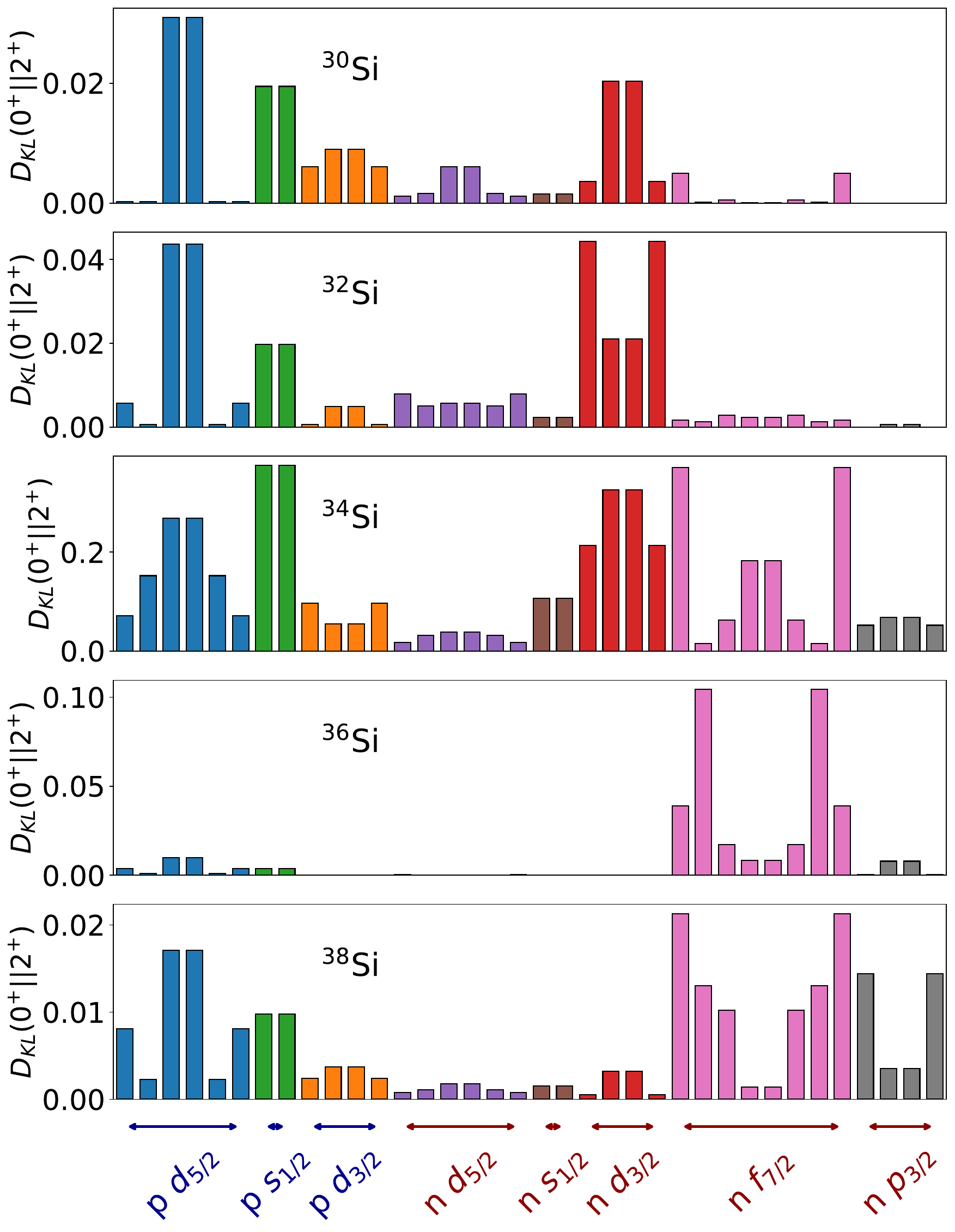}
        \label{fig:sub1}
    \end{subfigure}
    \hfill
    \begin{subfigure}[b]{0.49\textwidth}
        \includegraphics[width=\textwidth]{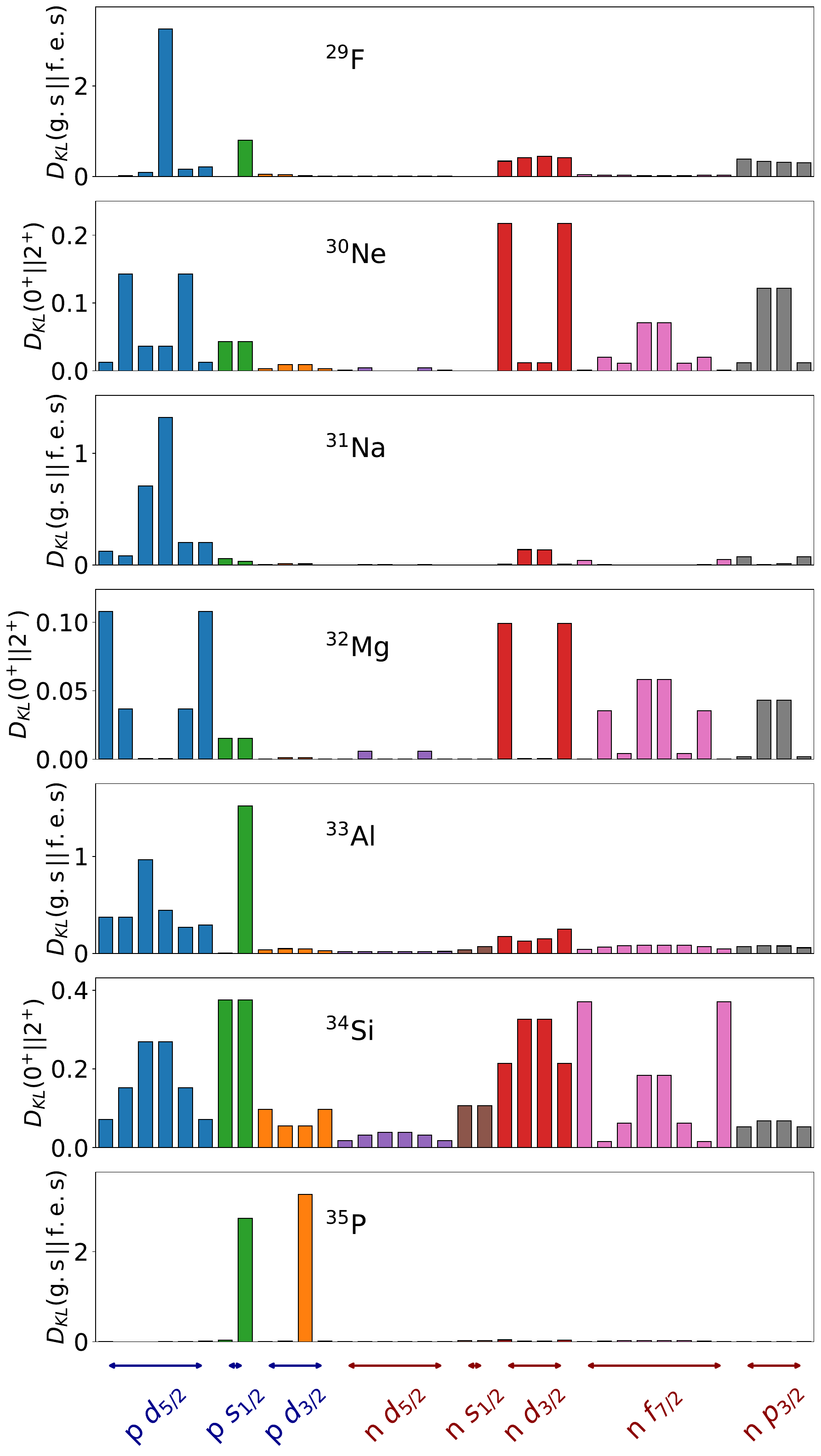}
        \label{fig:sub2}
    \end{subfigure}
    \caption{Left Panels: Same as Fig.\ref{fig:Mode_KLD_Ne_Mg}, but for $^{30-38}$Si isotopic chain.
    Right Panels: Same as Fig.\ref{fig:Mode_KLD_Ne_Mg}, but for the $^{29}$F to $^{35}$P isotonic chain. For even-odd nuclei, we compare the ground states and the first excited states.}
    \label{fig:Mode_KLD_Si_Isotone}
\end{figure*}

We now continue our analysis of the Mg isotopic chain. In the right column of Fig. \ref{fig:Mode_KLD_Ne_Mg}, we have presented the mode-resolved quantum relative entropy for the Mg isotopes. Similar to the Ne isotopic chain, here the proton $d_{5/2}$ orbitals also contribute the most to the quantum relative entropy. Among the neutron orbitals, ${}^{26}$Mg shows minimal contributions from the $d_{5/2}$, $s_{1/2}$, and $d_{3/2}$ orbitals. The ${}^{28}$Mg and ${}^{30}$Mg have indistinguishability only in $d_{3/2}$ orbitals. Although the quantum relative entropy for ${}^{34}$Mg and ${}^{36}$Mg is spanned across the $d_{3/2}$, $f_{7/2}$, and $p_{3/2}$ orbitals, the magnitude is very low compared to ${}^{32}$Mg. Now, we will analyze the right panel of Fig. \ref{fig:KLD_JSD_Ne_Mg}. The total mode-resolved quantum relative entropy increases as we move towards the IoI, peaking at ${}^{32}$Mg, and then decreases again. The proton-neutron quantum relative entropy follows a similar trend for both the KLD and JSD measures: the entropy dips at ${}^{28}$Mg, increases at ${}^{30}$Mg, and then falls for the following three isotopes. The quantum relative entropy for the Slater determinant has a dip at ${}^{30}$Mg, and is almost constant for all other isotopes, a trend followed by both relative entropies.

\begin{figure*}[t]
    \centering

    \begin{subfigure}[b]{0.49\textwidth}
        \includegraphics[width=\textwidth]{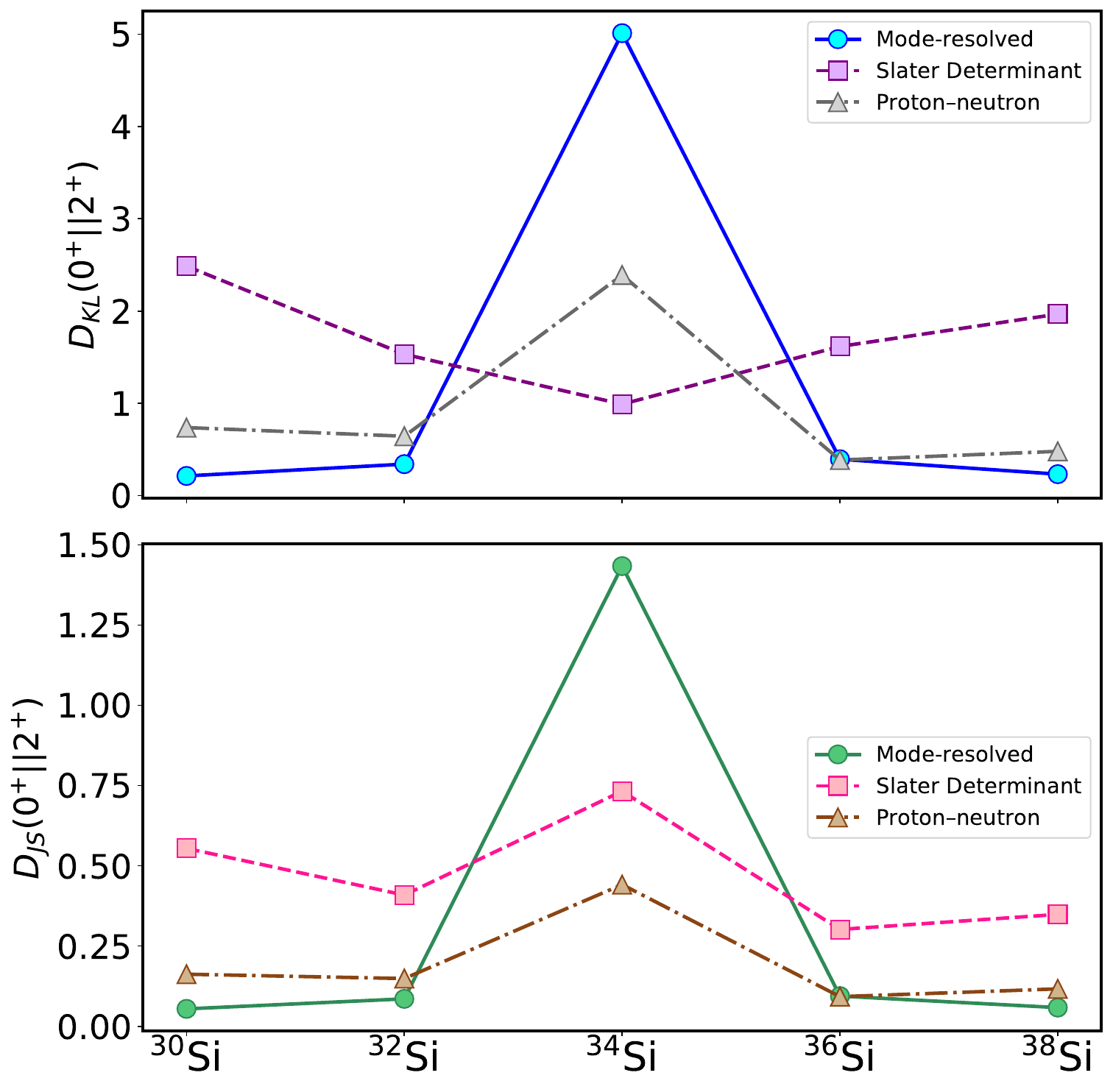}
        \label{fig:sub1}
    \end{subfigure}
    \hfill
    \begin{subfigure}[b]{0.48\textwidth}
        \includegraphics[width=\textwidth]{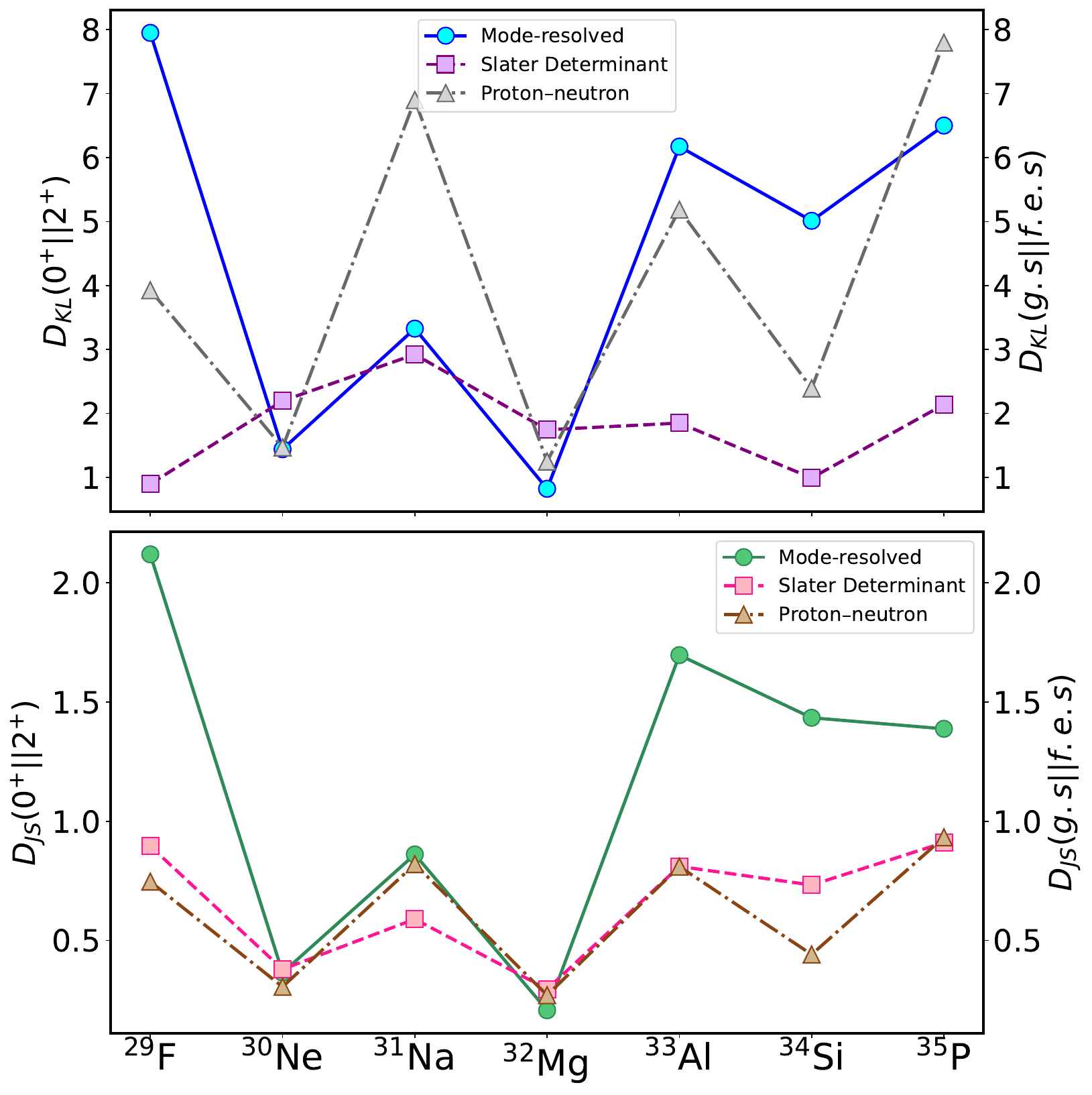}
        \label{fig:sub2}
    \end{subfigure}
    \caption{Same as for Fig.\ref{fig:KLD_JSD_Ne_Mg}, but for $^{30-38}$Si isotopic chain and $N=20$ isotonic chain.}
    \label{fig:KLD_JSD_Si_Isotone}
\end{figure*}

For the Si isotopes as shown in Fig.\ref{fig:Mode_KLD_Si_Isotone}, the magnitude of the mode-resolved quantum relative entropy remains very small for all nuclei except ${}^{34}$Si. The configurations of the $0^+$ and $2^+$ states are nearly identical across the isotopic chain, except in ${}^{34}$Si, where the $2^+$ state shows a significantly larger contribution from intruder configurations compared to the $0^+$ state, as shown in the Fig. \ref{fig:pn_ent_Ne_Mg_Si}. This leads to a strong distinguishability in ${}^{34}$Si, as seen in Fig.~\ref{fig:KLD_JSD_Si_Isotone}. This behavior is also reflected in the total mode-resolved relative entropy for both the KLD and JSD measures. The proton-neutron relative entropy follows a similar trend as before, though its peak is less pronounced than that of the mode-resolved relative entropy. For the Slater determinant partition, the KLD shows its lowest value at ${}^{34}$Si, increasing gradually for isotopes on either side. In contrast, the JSD reaches a maximum at ${}^{34}$Si, consistent with the behavior observed in the other two partitions.

The rightmost panels in Fig. \ref{fig:Mode_KLD_Si_Isotone} represent the mode-resolved quantum relative entropy for the $N=20$ isotones. For the odd-even nuclei, we have considered the relative entropy between the ground state and the first excited state. For ${}^{29}$F nucleus, major contribution to the distinguishability comes from the $m=1/2$ orbital of the proton $d_{5/2}$ orbitals and also some minor contribution from $m=1/2$ of $s_{1/2}$ proton orbitals, the neutron orbitals $d_{3/2}$ and $p_{3/2}$ has minor contributions as well. We have already discussed the ${}^{30}$Ne and ${}^{32}$Mg nuclei, and the contributions are very low when compared to the nuclei outside the IoI. Even ${}^{31}$Na has low relative entropy for all the orbitals, with the exception of $d_{5/2}$ proton orbitals. ${}^{33}$Al is distinguishable in the proton orbitals due to high relative entropy contributions from proton $d_{5/2}$ and $s_{1/2}$ orbitals. There are contributions from almost all orbitals in the ${}^{34}$Si, causing the total relative entropy to increase. For ${}^{35}$P, only $s_{1/2}$ and $d_{3/2}$ proton orbitals contribute to the relative entropy. Now we will discuss the total quantum relative entropies in the right panels of Fig. \ref{fig:KLD_JSD_Si_Isotone}. The total mode-resolved entropy for the $N=20$ shows some features related to the IoI, the entropies of ${}^{30}$Ne, ${}^{31}$Na, and ${}^{32}$Mg are lower compared to other nuclei, for these calculations we have considered the $0^+$ and $2^+$ states for even-even nuclei, and the ground and first excited states for odd-even nuclei. The low relative entropy for the IoI can be attributed to the indistinguishability between the states due to the collectivity in the states being compared. For ${}^{29}$F, ${}^{33}$Al, and ${}^{35}$P, the high relative entropy comes from the proton orbitals, whereas for ${}^{34}$Si, there are contributions from all orbitals. The proton-neutron relative entropy is lower for even-even nuclei and higher for odd-even nuclei; these patterns can be attributed to unpaired protons in the odd-even nuclei, causing the distinguishability to increase. The KLD has no discernible pattern for the Slater determinant partition, as it is not a true distance measure. However, the JSD shows a dip for the nuclei residing in the IoI, and the other nuclei have a slightly higher JSD, but the difference is not large when compared to the mode-resolved JSD.

\section{Summary and Conclusion} \label{summary}

In this work, we have carried out an extensive quantum information analysis of neutron-rich nuclei in and around the IoI using shell-model wavefunctions derived from VS-IMSRG. We have used various metrics to quantify entanglement and distinguishability between states, including proton-neutron entanglement entropy, mutual information, and quantum relative entropy. 
We have noticed from our studies that entanglement properties are sensitive to the structural evolutions around the IoI and are reflected in the metrics being calculated. 

The proton-neutron entanglement entropy was found to be a sensitive indicator of cross-shell excitations and the emergence of intruder states. In the Ne and Mg isotopic chains, a pronounced increase in entropy at $N=20$ shows the weakening of the shell gap and the onset of strong proton-neutron correlations associated with neutron excitations into the $pf$ shell. In contrast, the Si isotope exhibits significantly lower entanglement in its ground state at $N=20$, consistent with its location outside the IoI.

The orbital mutual information analysis provided a detailed, localized correlation, revealing dominant isovector pairing patterns in the like-particle sectors and systematically weaker, proton-neutron correlations. The excited states exhibit dominant isovector pairings, along with off-diagonal pairings in the like-particle sectors, and also display proton-neutron correlations with a strength comparable to that of the like-particle sector. Apart from the ground states of ${}^{28}$Ne and ${}^{30}$Mg showing high proton-neutron correlations in the $d_{3/2}$ orbitals, which hints towards the clumping of neutrons in the $sd$ shell, there are not many patterns observed in the proton-neutron sectors.

Quantum relative entropy provided a complementary perspective by quantifying the distinguishability between ground and excited states. Mode-resolved analyses demonstrated that structural changes across the IoI are driven primarily by neutron orbitals associated with cross-shell excitations. The $N=20$ isotone chain results for mode-resolved JSD and KLD suggest that the distinguishability is low for the IoI nuclei as compared to the outlying nuclei.

Since the $sdpf$ model space consists of two major oscillator shells, the dimensionality of the many-body Hilbert space increases exponentially with nucleon number. Studying entanglement properties in this region may therefore offer valuable insights for optimizing and developing efficient simulation strategies for high-dimensional many-body systems, including possible applications on quantum computing platforms. The proton-neutron entanglement entropy provides us with the possibility to perform independent simulations for the proton and neutron model spaces with low entanglement entropy \cite{CJ_2024}. Furthermore, quantum relative entropy, as a recently introduced tool in nuclear structure studies, may open new avenues for exploring connections between nuclear correlations, state distinguishability, and thermodynamic concepts in finite nuclei. There is also a possibility that quantum JSD can serve as a robust optimization and diagnostic measure in quantum simulations.

\section*{Acknowledgement}
We acknowledge financial support from 
MHRD (India) and research Grant No.
ANRF/ARGM/2025/001130/TS from Anusandhan National Research Foundation [ANRF], India.
We would also like to thank the National Supercomputing Mission (NSM) for providing computing resources of ‘PARAM Ganga’ at
the Indian Institute of Technology Roorkee.
We are grateful to Calvin Johnson for his help regarding the BIGSTICK and to Takayuki Miyagi for IM-SRG.


\begin{thebibliography}{1}
	
	\expandafter\ifx\csname natexlab\endcsname\relax\def\natexlab#1{#1}\fi
	\expandafter\ifx\csname bibnamefont\endcsname\relax
	\def\bibnamefont#1{#1}\fi
	\expandafter\ifx\csname bibfnamefont\endcsname\relax
	\def\bibfnamefont#1{#1}\fi
	\expandafter\ifx\csname citenamefont\endcsname\relax
	\def\citenamefont#1{#1}\fi
	\expandafter\ifx\csname url\endcsname\relax
	\def\url#1{\texttt{#1}}\fi
	\expandafter\ifx\csname urlprefix\endcsname\relax\def\urlprefix{URL }\fi
	\providecommand{\bibinfo}[2]{#2}
	\providecommand{\eprint}[2][]{\url{#2}}
	


\bibitem{chuang}
M. A. Nielsen and I. L. Chuang, Quantum Computation and
Quantum Information: 10th Anniversary Edition (Cambridge University Press, Cambridge, England, 2010).

\bibitem{nisq} J. Preskill, Quantum Computing in the NISQ era and beyond, \href{https://doi.org/10.22331/q-2018-08-06-79}{Quantum {\bf 2}, 79 (2018).}

\bibitem{vqe} A. Peruzzo, J. McClean, P. Shadbolt et al. A variational eigenvalue solver on a photonic quantum processor, \href{https://doi.org/10.1038/ncomms5213}{Nat. Commun. 5, 4213 (2014).}

\bibitem{Perez2_2023}
A. Pérez‑Obiol, A. M. Romero, J. Menéndez, A. Rios, A. García‑Sáez, B. Juliá‑Díaz,
Nuclear shell‑model simulation in digital quantum computers,
\href{https://doi.org/10.1038/s41598-023-39263-7}{Sci. Rep. \textbf{13}, 12291 (2023).}

\bibitem{Lacroix_PRL}
 D. Lacroix, Symmetry-assisted preparation of entangled many-body states on a quantum computer, 
 \href{https://doi.org/10.1103/PhysRevLett.125.230502}
 {Phys. Rev. Lett. {\bf 125}, 230502 (2020).}

\bibitem{Ayral_Besserve_Lacroix_2023}
T. Ayral, P. Besserve, D. Lacroix, E. A. R. Guzman,
Quantum computing with and for many-body physics,
\href{https://doi.org/10.1140/epja/s10050-023-01141-1}{Eur. Phys. J. A \textbf{59}, 227 (2023).}

\bibitem{Chandan_2023}
C. Sarma, O. D. Matteo, A. Abhishek, and P. C. Srivastava,
Prediction of the neutron drip line in oxygen isotopes using quantum computation,
\href{https://doi.org/10.1103/PhysRevC.108.064305}{Phys. Rev. C {\bf 108}, 064305 (2023).}

\bibitem{Bharti_Bhoy}
 B. Bhoy, P. Stevenson, 
 Shell-model study of $^{58}$Ni using quantum computing algorithm,
\href{https://doi.org/10.1088/1367-2630/ad5756} {New J. Phys. {\bf 26}, 075001 (2024).}
 

\bibitem{QC_n1} E. F. Dumitrescu, A. J. McCaskey, G. Hagen, G. R.
Jansen, T. D. Morris, T. Papenbrock, R. C. Pooser, D. J.
Dean, and P. Lougovski, Cloud Quantum Computing of an Atomic Nucleus, \href{https://doi.org/10.1103/PhysRevLett.120.210501}{Phys. Rev. Lett. 120, 210501 (2018).}

\bibitem{QC_n2} H.-H. Lu, N. Klco, J. M. Lukens, T. D. Morris, A. Bansal,
A. Ekström, G. Hagen, T. Papenbrock, A. M. Weiner,
M. J. Savage, and P. Lougovski, Simulations of subatomic many-body physics on a quantum frequency processor, \href{https://doi.org/10.1103/PhysRevA.100.012320}{Phys. Rev. A {\bf 100}, 012320 (2019).}

\bibitem{graycode} O. Di Matteo, A. McCoy, P. Gysbers, T. Miyagi, R. M. Woloshyn, and P. Navrátil, Improving Hamiltonian encodings with the Gray code, \href{https://doi.org/10.1103/PhysRevA.103.042405}{Phys. Rev. A 103, 042405 (2021).}





\bibitem{pooja2} P. Siwach and P. Arumugam, Quantum computation of nuclear observables involving linear combinations of unitary operators, \href{https://doi.org/10.1103/PhysRevC.105.064318}{Phys. Rev. C {\bf 105}, 064318 (2022).}


\bibitem{QC_n3} I. Stetcu, A. Baroni, and J. Carlson, Variational approaches to constructing the many-body nuclear ground state for quantum computing, \href{https://doi.org/10.1103/PhysRevC.105.064308}{Phys. Rev. C {\bf 105}, 064308 (2022).}

\bibitem{QC_n4} A. M. Romero, J. Engel, H. L. Tang, and S. E.
Economou, Solving nuclear structure problems with the adaptive variational quantum algorithm, \href{https://doi.org/10.1103/PhysRevC.105.064317}{ Phys. Rev. C {\bf 105}, 064317 (2022).}

\bibitem{QC_n5} O. Kiss, M. Grossi, P. Lougovski, F. Sanchez, S. Vallecorsa, and T. Papenbrock, Quantum computing of the $^6$Li nucleus via ordered unitary coupled clusters, \href{https://doi.org/10.1103/PhysRevC.106.034325}{Phys. Rev. C {\bf 106}, 034325 (2022).}

\bibitem{Robin_Savage_2023}
C. E. P. Robin and M. J. Savage,
Quantum simulations in effective model spaces: Hamiltonian-learning variational quantum eigensolver using digital quantum computers and application to the Lipkin–Meshkov–Glick model,
\href{https://doi.org/10.1103/PhysRevC.108.024313}{Phys. Rev. C \textbf{108}, 024313 (2023).}


\bibitem{Denis_2024}
J. Zhang, D. Lacroix, and Y. Beaujeault-Taudière, 
Neutron-proton pairing correlations described on quantum computers,
\href{https://doi.org/10.1103/PhysRevC.110.064320}{Phys. Rev. C \textbf{110}, 064320 (2024).}


\bibitem{Yoshida_2024}
S. Yoshida, T. Sato, T. Ogata, T. Naito, and M. Kimura,
Accurate and precise quantum computation of valence two-neutron systems,
\href{https://doi.org/10.1103/PhysRevC.109.064305}{Phys. Rev. C \textbf{109}, 064305 (2024).}

\bibitem{Dean}
D. Lee, Quantum techniques for eigenvalue problems, 
\href{https://doi.org/10.1140/epja/s10050-023-01183-5}{Eur. Phys. J. A \textbf{59}, 275 (2023).}


\bibitem{Vary_2021}
W. Du, J. P. Vary, X. Zhao, and W. Zuo,
Quantum simulation of nuclear inelastic scattering,
\href{https://doi.org/10.1103/PhysRevA.104.012611}{Phys. Rev. A \textbf{104}, 012611 (2021).}






\bibitem{Horodecki_2009}
R. Horodecki, P. Horodecki, M. Horodecki, and K. Horodecki,
Quantum entanglement,
\href{https://doi.org/10.1103/RevModPhys.81.865}{Rev. Mod. Phys. \textbf{81}, 865 (2009).}

\bibitem{Hill_Wootters_1997}
S. Hill and W. K. Wootters,
Entanglement of a Pair of Quantum Bits,
\href{https://doi.org/10.1103/PhysRevLett.78.5022}{Phys. Rev. Lett. \textbf{78}, 5022 (1997).}

\bibitem{Belavkin_Ohya_2002}
V. P. Belavkin and M. Ohya,
Entanglement, quantum entropy and mutual information,
\href{https://doi.org/10.1098/rspa.2001.0867}{Proc. R. Soc. A \textbf{458}, 209 (2002).}

\bibitem{Eisert_2006}
J. Eisert,
Entanglement in quantum information theory,
\href{https://doi.org/10.48550/arXiv.quant-ph/0610253}{arXiv:quant-ph/0610253 (2006).}

\bibitem{Gigena_Rossignoli_2015}
N. Gigena and R. Rossignoli,
Entanglement in fermion systems,
\href{https://doi.org/10.1103/PhysRevA.92.042326}{Phys. Rev. A \textbf{92}, 042326 (2015).}

\bibitem{Schliemann_2001}
J. Schliemann, J. I. Cirac, M. Kuś, M. Lewenstein, and D. Loss,
Quantum correlations in two-fermion systems,
\href{https://doi.org/10.1103/PhysRevA.64.022303}{Phys. Rev. A \textbf{64}, 022303 (2001).}

\bibitem{Eckert_2002}
K. Eckert, J. Schliemann, D. Bruß, and M. Lewenstein,
Quantum correlations in systems of indistinguishable particles,
\href{https://doi.org/10.1006/aphy.2002.6268}{Ann. Phys. \textbf{299}, 88 (2002).}

\bibitem{Wiseman_Vaccaro_2003}
H. M. Wiseman and J. A. Vaccaro,
Entanglement of indistinguishable particles shared between two parties,
\href{https://doi.org/10.1103/PhysRevLett.91.097902}{Phys. Rev. Lett. \textbf{91}, 097902 (2003).}

\bibitem{Ghirardi_Marinatto_2004}
G. C. Ghirardi and L. Marinatto,
General criterion for the entanglement of two indistinguishable particles,
\href{https://doi.org/10.1103/PhysRevA.70.012109}{Phys. Rev. A \textbf{70}, 012109 (2004).}


\bibitem{Benatti_Floreanini_Titimbo_2014}
F. Benatti, R. Floreanini, and K. Titimbo,
Entanglement of identical particles,
\href{https://doi.org/10.1142/S1230161214400034}{Open Syst. Inf. Dyn. \textbf{21}, 1440003 (2014).}

\bibitem{LoFranco_Compagno_2016}
R. Lo Franco and G. Compagno,
Quantum entanglement of identical particles by standard information-theoretic notions,
\href{https://doi.org/10.1038/srep20603}{Sci. Rep. \textbf{6}, 20603 (2016).}


\bibitem{Benatti_Floreanini_Marzolino_2014}
F. Benatti, R. Floreanini, and U. Marzolino,
Entanglement in fermion systems and quantum metrology,
\href{https://doi.org/10.1103/PhysRevA.89.032326}{Phys. Rev. A \textbf{89}, 032326 (2014).}

\bibitem{Zanardi_2002}
P. Zanardi,
Quantum entanglement in fermionic lattices,
\href{https://doi.org/10.1103/PhysRevA.65.042101}{Phys. Rev. A \textbf{65}, 042101 (2002).}

\bibitem{Shi_2003}
Y. Shi,
Quantum entanglement of identical particles,
\href{https://doi.org/10.1103/PhysRevA.67.024301}{Phys. Rev. A \textbf{67}, 024301 (2003).}

\bibitem{Friis_2013}
N. Friis, A. R. Lee, and D. E. Bruschi,
Fermionic-mode entanglement in quantum information,
\href{https://doi.org/10.1103/PhysRevA.87.022338}{Phys. Rev. A \textbf{87}, 022338 (2013).}

\bibitem{Coffman_Kundu_Wootters_2000}
V. Coffman, J. Kundu, and W. K. Wootters,
Distributed entanglement,
\href{https://doi.org/10.1103/PhysRevA.61.052306}{Phys. Rev. A \textbf{61}, 052306 (2000).}

\bibitem{Wong_Christensen_2001}
A. Wong and N. Christensen,
Potential multiparticle entanglement measure,
\href{https://doi.org/10.1103/PhysRevA.63.044301}{Phys. Rev. A \textbf{63}, 044301 (2001).}


\bibitem{Baid_2024}
S. Baid, \textit{et al.},
Extended Lipkin model: Proposal for implementation in a quantum platform and machine learning analysis of its phase diagram,
\href{https://doi.org/10.1103/PhysRevC.110.044318}{Phys. Rev. C \textbf{110}, 044318 (2024).} 

\bibitem{Hengstenberg_2023}
S. M. Hengstenberg, C. E. P. Robin, and M. J. Savage,
Multi-body entanglement and information rearrangement in nuclear many-body systems: a study of the Lipkin–Meshkov–Glick model,
\href{https://doi.org/10.1140/epja/s10050-023-01145-x}{Eur. Phys. J. A \textbf{59}, 231 (2023).}

\bibitem{Latorre_2005}
J. I. Latorre, R. Orús, E. Rico, and J. Vidal,
Entanglement entropy in the Lipkin–Meshkov–Glick model,
\href{https://doi.org/10.1103/PhysRevA.71.064101}{Phys. Rev. A \textbf{71}, 064101 (2005).}

\bibitem{Vidal_Palacios_Aslangul_2004}
J. Vidal, G. Palacios, and C. Aslangul,
Entanglement dynamics in the Lipkin–Meshkov–Glick model,
\href{https://doi.org/10.1103/PhysRevA.70.062304}{Phys. Rev. A \textbf{70}, 062304 (2004).}

\bibitem{Carl_2021}
C. Robin, M. J. Savage, and N. Pillet,
Entanglement rearrangement in self-consistent nuclear structure calculations,
\href{https://doi.org/10.1103/PhysRevC.103.034325}{Phys. Rev. C \textbf{103}, 034325 (2021).}

\bibitem{Sarma_2024}
C. Sarma and P. C. Srivastava,
Investigation of entanglement in $N = Z$ nuclei within no-core shell model,
\href{https://doi.org/10.1016/j.nuclphysa.2026.123339}{Nucl. Phys. A \textbf{1068}, 123339 (2026).} 


\bibitem{Kruppa_2022}
A. T. Kruppa, J. Kovacs, P. Salamon, O. Legeza, G. Zarand,
Entanglement and seniority,
\href{https://doi.org/10.1103/PhysRevC.106.024303}{Phys. Rev. C \textbf{106}, 024303 (2022).}

\bibitem{Tichai_2023}
A. Tichai, S. Knecht, A.T. Kruppa, Ö. Legeza, C.P. Moca, A. Schwenk, M.A. Werner, G. Zarand,
Combining the in-medium similarity renormalization group with the density matrix renormalization group: Shell structure and information entropy,
\href{https://doi.org/10.1016/j.physletb.2023.138139}{Phys. Lett. B \textbf{845} 138139 (2023).}

\bibitem{Pazy_2023}
E. Pazy,
Entanglement entropy between short-range correlations and the Fermi sea in nuclear structure,
\href{https://doi.org/10.1103/PhysRevC.107.054308}{Phys. Rev. C \textbf{107}, 054308 (2023).}

\bibitem{Liang_2025}
S. Y. Liang, Y. Lu, Y. Lei, C. W. Johnson, G. J. Fu, and J. J. Shen,
Shannon entropy of optimized proton-neutron pair condensates,
\href{https://doi.org/10.1103/PhysRevC.111.024310}{Phys. Rev. C \textbf{111}, 024310 (2025).}

\bibitem{Gu_2023}
C. Gu, Z. H. Sun, G. Hagen, and T. Papenbrock,
Entanglement entropy of nuclear systems,
\href{https://doi.org/10.1103/PhysRevC.108.054309}{Phys. Rev. C \textbf{108}, 054309 (2023).}

\bibitem{Bai_Ren_2022}
D. Bai and Z. Ren,
Entanglement generation in few-nucleon scattering,
\href{https://doi.org/10.1103/PhysRevC.106.064005}{Phys. Rev. C \textbf{106}, 064005 (2022).}

\bibitem{Bai_2023}
D. Bai,
Spin entanglement in neutron-proton scattering,
\href{https://doi.org/10.1016/j.physletb.2023.138162}{Phys. Lett. B \textbf{845}, 138162 (2023).}

\bibitem{Carl_2025}
F. Brökemeier, S. M. Hengstenberg, J. W. T. Keeble, C. Robin, F. Rocco, M. J. Savage,
Quantum magic and multipartite entanglement in the structure of nuclei,
\href{https://doi.org/10.1103/PhysRevC.111.034317}{Phys. Rev. C \textbf{111}, 034317 (2025).}

\bibitem{CJ_2023}
C. W. Johnson, O. C. Gorton,
Proton-neutron entanglement in the nuclear shell model, 
\href{https://doi.org/10.1088/1361-6471/acbece}{J. Phys. G: Nucl. Part. Phys. \textbf{50}, 045110 (2023).}

\bibitem{Perez_2023}
A. Pérez-Obiol, S. Masot-Llima, A. M. Romero, \textit{et al.},
Quantum entanglement patterns in the structure of atomic nuclei within the nuclear shell model,
\href{https://doi.org/10.1140/epja/s10050-023-01151-z}{Eur. Phys. J. A \textbf{59}, 240 (2023).}

\bibitem{Kruppa_2021}
A. T. Kruppa, J. Kovacs, P. Salamon, and O. Legeza,
Entanglement and correlation in two-nucleon systems,
\href{https://doi.org/10.1088/1361-6471/abc2dd}{J. Phys. G: Nucl. Part. Phys. \textbf{48}, 025107 (2021).}

\bibitem{Kruppa_2025}
J. Kovacs, A. T. Kruppa, O. Legeza, and P. Salamon,
Mode entanglement and isospin pairing in two-nucleon systems,
\href{https://doi.org/10.1088/1361-6471/ad9345}{J. Phys. G: Nucl. Part. Phys. \textbf{52}, 015105 (2025).}


\bibitem{PerezObiol_2024}
A. Pérez-Obiol, S. Masot-Llima, A. M. Romero, J. Menéndez, A. Rios, A. García-Sáez, and B. Juliá-Díaz,
Entropy-driven entanglement forging,
\href{https://doi.org/10.48550/arXiv.2409.04510}{arXiv:2409.04510 [quant-ph] (2024).}


\bibitem{Shinde_2025}
R. M. Shinde and P. C. Srivastava,
Study of entanglement in Ne, Mg, and Si isotopic chains,
\href{https://doi.org/10.1016/j.jspc.2025.100142}{J. Subat. Part. Cosmology \textbf{4}, 100142 (2025).}

\bibitem{Costa_2025}
E. Costa, A. Pérez-Obiol, J. Menéndez, A. Rios, A. García-Sáez, and B. Juliá-Díaz,
A quantum annealing protocol to solve the nuclear shell model,
\href{https://doi.org/10.21468/SciPostPhys.19.2.062}{SciPost Phys. \textbf{19}, 062 (2025).}

\bibitem{Johnson_2025}
C. W. Johnson, O. C. Gorton,
A weak entanglement approximation for nuclear structure: a progress report, \href{https://doi.org/10.1016/j.jspc.2025.100061}{J. Subat. Part. Cosmology \textbf{3}, 100061 (2025).}

\bibitem{CJ_2024}
O. C. Gorton, C. W. Johnson, 
Weak entanglement approximation for nuclear structure, 
\href{https://doi.org/10.1103/PhysRevC.110.034305}{Phys. Rev. C \textbf{110}, 034305 (2024).}


\bibitem{Vedral_2002}
V. Vedral,
The role of relative entropy in quantum information theory,
\href{https://doi.org/10.1103/RevModPhys.74.197}{Rev. Mod. Phys. \textbf{74}, 197 (2002).}

\bibitem{Schumacher_Westmoreland_2000}
B. Schumacher and M. D. Westmoreland,
Relative entropy in quantum information theory,
\href{https://doi.org/10.48550/arXiv.quant-ph/0004045}{arXiv:quant-ph/0004045 (2000).}

\bibitem{Sagawa_2012}
T. Sagawa,
Second Law-Like Inequalities with Quantum Relative Entropy: An Introduction,
\href{https://doi.org/10.48550/arXiv.1202.0983}{arXiv:1202.0983 [quant-ph](2012).} 


\bibitem{Floerchinger_2020}
S. Floerchinger and T. Haas,
Thermodynamics from relative entropy,
\href{https://doi.org/10.1103/PhysRevE.102.052117}{Phys. Rev. E \textbf{102}, 052117 (2020).}

\bibitem{Majtey_2005}
A. P. Majtey, P. W. Lamberti, and D. P. Prato,
Jensen-Shannon divergence as a measure of distinguishability between mixed quantum states,
\href{https://doi.org/10.1103/PhysRevA.72.052310}{Phys. Rev. A \textbf{72}, 052310 (2005).}

\bibitem{Grosse_2002}
I. Grosse, P. Bernaola-Galv{\'a}n, P. Carpena, R. Rom{\'a}n-Rold{\'a}n, J. Oliver, and H. E. Stanley,
Analysis of symbolic sequences using the Jensen-Shannon divergence,
\href{https://doi.org/10.1103/PhysRevE.65.041905}{Phys. Rev. E \textbf{65}, 041905 (2002).}

\bibitem{Lamberti_2008}
P. W. Lamberti, A. P. Majtey, A. Borr{\'a}s, M. Casas, and A. Plastino,
Metric character of the quantum Jensen-Shannon divergence,
\href{https://doi.org/10.1103/PhysRevA.77.052311}{Phys. Rev. A \textbf{77}, 052311 (2008).}

\bibitem{Briet_2009}
J. Bri{\"e}t and P. Harremo{\"e}s,
Properties of classical and quantum Jensen-Shannon divergence,
\href{https://doi.org/10.1103/PhysRevA.79.052311}{Phys. Rev. A \textbf{79}, 052311 (2009).}



\bibitem{Caurier_2005}E. Caurier, G. Martinez-Pinedo, F. Nowacki, A. Poves, A. P. Zuker,
 The shell model as a unified view of nuclear structure,
\href{https://doi.org/10.1103/RevModPhys.77.427}
{Rev. of Mod. Phys.  {\bf 77}, 427-488 (2005).}

\bibitem{BrownReview}
B. A. Brown, The Nuclear Shell Model towards the Drip Lines,
\href{https://doi.org/10.3390/physics4020035}
{Physics {\bf 4}, 525 (2022).}

\bibitem{Otsuka_2020}
T. Otsuka, A. Gade, O. Sorlin, T. Suzuki, and Y. Utsuno,
Evolution of shell structure in exotic nuclei,
\href{https://doi.org/10.1103/RevModPhys.92.015002}{Rev. Mod. Phys. \textbf{92}, 015002 (2020).}

\bibitem{SorlinReview}
O. Sorlin, M.-G. Porquet, Nuclear magic numbers: New features far from stability,
\href{https://doi.org/10.1016/j.ppnp.2008.05.001}
{Prog. Part. Nucl. Phys. {\bf 61}, 602 (2008).}

\bibitem{WarburtonIoI}
E. K. Warburton, J. A. Becker, and B. A. Brown, Mass systematics for $A=29$-44 nuclei: The deformed $A \sim $32 region,
\href{https://doi.org/10.1103/PhysRevC.41.1147}
{Phys. Rev. C {\bf 41}, 1147 (1990).}

\bibitem{Poves1994}
A. Poves, J. Retamosa, Theoretical study of the very neutron-rich nuclei around $N=20$,
\href{https://doi.org/10.1016/0375-9474(94)90058-2}
{Nucl. Phys. A {\bf 571}, 221 (1994).}


\bibitem{Kaneko_2011}K. Kaneko, Y. Sun, T. Mizusaki, and M. Hasegawa,
 Shell-model study for neutron-rich sd-shell nuclei,
\href{https://doi.org/10.1103/PhysRevC.83.014320}
{Phys. Rev. C \textbf{83}, 014320 (2011).}



\bibitem{MCSM_1999}
Y. Utsuno, T. Otsuka, T. Mizusaki, and M. Honma, Varying shell gap and deformation in $N \sim 20$ unstable nuclei studied by the Monte Carlo shell model,
\href{https://doi.org/10.1103/PhysRevC.60.054315}
{Phys. Rev. C {\bf 60}, 054315 (1999).}


\bibitem{SDPF-U}
F. Nowacki and A. Poves, New effective interaction for $0\hbar \omega$ shell-model calculations in the $sd-pf$ valence space,
\href{https://doi.org/10.1103/PhysRevC.79.014310}
{Phys. Rev. C {\bf 79}, 014310 (2009).}

\bibitem{SDPF-MU}
Y. Utsuno, T. Otsuka, B. A. Brown, M. Honma, T. Mizusaki, and N. Shimizu, Shape transitions in exotic Si and S isotopes and tensor-force-driven Jahn-Teller effect,
\href{http://dx.doi.org/10.1103/PhysRevC.86.051301}
{Phys. Rev. C {\bf 86}, 051301(R) (2012).}

\bibitem{SDPF-U-MIX}
E. Caurier, F. Nowacki, and A. Poves, Merging of the islands of inversion at $N = 20$ and $N = 28$,
\href{https://doi.org/10.1103/PhysRevC.90.014302}
{Phys. Rev. C {\bf 90}, 014302 (2014).}


\bibitem{EEdf1}
N. Tsunoda, T. Otsuka, N. Shimizu, M. H. Jensen, K. Takayanagi, and T. Suzuki Exotic neutron-rich medium-mass nuclei with realistic nuclear forces,
\href{http://dx.doi.org/10.1103/PhysRevC.95.021304}
{Phys. Rev. C {\bf 95}, 021304(R) (2017).}






\bibitem{RagnarReview}
S. R. Stroberg, H. Hergert, S. K. Bogner, and J. D. Holt, Nonempirical Interactions for the Nuclear Shell Model: An Update,
\href{https://doi.org/10.1146/annurev-nucl-101917-021120}
{Annu. Rev. Nucl. Part. Sci. {\bf 69}, 307 (2019).}

\bibitem{HergertReview}
H. Hergert, S. K. Bogner, T. D. Morris, A. Schwenk, K. Tsukiyama, The In-Medium Similarity Renormalization Group: A Novel \textit{Ab Initio} Method for Nuclei,
\href{https://doi.org/10.1016/j.physrep.2015.12.007}
{Phys. Reports {\bf 621}, 165 (2016).}





\bibitem{Li2023PLB}
J.G. Li, H.H. Li, S. Zhang, Y.M. Xing, and W. Zuo,
Double-magicity of proton drip-line nucleus $^{22}$Si with ab initio calculation,
\href{https://doi.org/10.1016/j.physletb.2023.138197}
{Phys. Lett. B {\bf 846}, 138197 (2023).}

\bibitem{IMSRG_Magnus}
T. D. Morris, N. M. Parzuchowski, and S. K. Bogner, Magnus expansion and in-medium similarity renormalization group,
\href{http://dx.doi.org/10.1103/PhysRevC.92.034331}
{Phys. Rev. C {\bf 92}, 034331 (2015).}

\bibitem{IMSRG3f2}
B. C. He and S. R. Stroberg, Factorized approximation to the in-medium similarity renormalization group IMSRG(3),
\href{https://doi.org/10.1103/PhysRevC.110.044317}
{Phys. Rev. C {\bf 110}, 044317 (2024).}

\bibitem{IMSRG3N7_Heinz2021}
M. Heinz, A. Tichai, J. Hoppe, K. Hebeler, and A. Schwenk, In-medium similarity renormalization group with three-body operators,
\href{https://doi.org/10.1103/PhysRevC.103.044318}
{Phys. Rev. C {\bf 103}, 044318 (2021).}


\bibitem{IMSRG3N7_Ragnar}
S. R. Stroberg, T. D. Morris, and B. C. He, In-medium similarity renormalization group with flowing 3-body operators, and approximations thereof,
\href{https://doi.org/10.1103/PhysRevC.110.044316}
{Phys. Rev. C {\bf 110}, 044316 (2024).}

\bibitem{IMSRG3N7_Heinz2024}
M. Heinz, T. Miyagi, S. R. Stroberg, A. Tichai, K. Hebeler, and A. Schwenk, Improved structure of calcium isotopes from \textit{ab initio} calculations,
\href{https://doi.org/10.1103/PhysRevC.111.034311}
{Phys. Rev. C {\bf 111}, 034311 (2025).}



\bibitem{IMSRG_RagnarPRL}
S. R. Stroberg, A. Calci, H. Hergert, J. D. Holt, S. K. Bogner, R. Roth, and A. Schwenk, Nucleus-Dependent Valence-Space Approach to Nuclear Structure,
\href{http://dx.doi.org/10.1103/PhysRevLett.118.032502}
{Phys. Rev. Lett. {\bf 118}, 032502 (2017).}


\bibitem{IMSRG_RagnarPRC}
S. R. Stroberg, H. Hergert, J. D. Holt, S. K. Bogner, and A. Schwenk,
Ground and excited states of doubly open-shell nuclei from ab initio valence-space Hamiltonians,
\href{http://dx.doi.org/10.1103/PhysRevC.93.051301}
{Phys. Rev. C {\bf 93}, 051301(R) (2016).}

\bibitem{Miyagi2022PRC}
T. Miyagi, S. R. Stroberg, P. Navrátil, K. Hebeler, and J. D. Holt, Converged \textit{ab initio} calculations of heavy nuclei,
\href{https://doi.org/10.1103/PhysRevC.105.014302}
{Phys. Rev. C {\bf 105}, 014302 (2022).}



\bibitem{IMSRG_Miyagi}
T. Miyagi, S. R. Stroberg, J. D. Holt, and N. Shimizu, Ab initio multishell valence-space Hamiltonians and the island of inversion,
\href{http://dx.doi.org/10.1103/PhysRevC.102.034320}
{Phys. Rev. C {\bf 102}, 034320 (2020).}

\bibitem{Yuan_Hu_2024}
Q. Yuan and B. S. Hu,
\textit{Ab initio} calculations of anomalous seniority breaking in the $\pi g_{9/2}$ shell for the $N=50$ isotones,
\href{https://doi.org/10.1016/j.physletb.2024.139018}{Phys. Lett. B \textbf{858}, 139018 (2024).}

\bibitem{Subhrajit1}
S. Sahoo and P.C. Srivastava, Evolution of shell structure at N=32 and 34: Insights from realistic nuclear forces,
\href{https://doi.org/10.1103/423y-znv8}
{Phys. Rev. C  {\bf 112}, L021301 (2025).} 


\bibitem{Subhrajit2}
S. Sahoo and P.C. Srivastava, \textit{Ab initio} study of the island of inversion in odd-A nuclei: Structure of $^{31,33}$Mg,
\href{https://doi.org/10.1103/PhysRevC.111.054308}
{Phys. Rev. C  {\bf 111}, 054308 (2025).} 

\bibitem{Simonis_2016}
J. Simonis, K. Hebeler, J. D. Holt, J. Menéndez, and A. Schwenk,
Exploring sd-shell nuclei from two- and three-nucleon interactions with realistic saturation properties,
\href{https://doi.org/10.1103/PhysRevC.93.011302}{Phys. Rev. C \textbf{93}, 011302(R) (2016).}

\bibitem{Hebeler_2011}
K. Hebeler, S. K. Bogner, R. J. Furnstahl, A. Nogga, and A. Schwenk,
Improved nuclear matter calculations from chiral low-momentum interactions,
\href{https://doi.org/10.1103/PhysRevC.83.031301}{Phys. Rev. C \textbf{83}, 031301(R) (2011).}



\bibitem{Johnson_2018}
C. W. Johnson, W. E. Ormand, K. S. McElvain, H. Shan,
Big-stick: A flexible configuration-interaction shell-model code, 
\href{https://arxiv.org/abs/1801.08432}{arXiv preprint arXiv:1801.08432 (2018).}





\bibitem{Praveen}
P.C. Srivastava and V. Kumar,
Spectroscopic factor strengths using \textit{ab initio} approaches,
\href{https://doi.org/10.1103/PhysRevC.94.064306} 
{Phys. Rev. {\bf C 94}, 064306 (2016).}

\bibitem{Na_work_NPA}
S. Sahoo, P.C. Srivastava, and T. Suzuki, Study of structure
and radii for $^{20-31}$Na isotopes using microscopic interactions,
\href{https://doi.org/10.1016/j.nuclphysa.2023.122618}
{Nucl. Phys. A {\bf 1032}, 122618 (2023).}


\bibitem{Priyanka1}
P. Choudhary, P.C. Srivastava, M. Gennari, P. Navratil,
\textit{Ab initio} no-core shell model description of $^{10-14}$C isotopes,
\href{https://doi.org/10.1103/PhysRevC.107.014309}
{Phys. Rev. C {\bf 107}, 014309  (2023).}

\bibitem{Priyanka2}
P. Choudhary, P.C. Srivastava, P. Navratil,
\textit{Ab initio} no-core shell model study of $^{10-14}$B isotopes with realistic $NN$ interactions,
\href{https://doi.org/10.1103/PhysRevC.102.044309} 
{Phys. Rev. {\bf C 102}, 044309 (2020).}

\bibitem{Chandan1}
C. Sarma and P.C. Srivastava,
\textit{Ab initio} no-core shell model study of $^{18-24}$Ne isotopes,
\href{https://doi.org/10.1088/1361-6471/acb962} 
{Journal of Physics G: Nuclear and Particle Physics  {\bf 50}, 045105 (2023).}




\bibitem{Suhonen}
J. Suhonen,
\textit{From Nucleons to Nucleus: Concepts of Microscopic Nuclear Theory} (Springer, Berlin, 2007).
\bibitem{Johnson_2013}
C. W. Johnson, W. E. Ormand, and P. G. Krastev,
Factorization in large-scale many-body calculations,
\href{https://doi.org/10.1016/j.cpc.2013.07.022}{Computer Phys. Commun. \textbf{184}, 2761–2774 (2013).}


\bibitem{Boguslawski_2013}
K. Boguslawski, 
P. Tecmer, G. Barcza, Ö. Legeza, and M. Reiher,
Orbital entanglement in bond-formation processes,
\href{https://doi.org/10.1021/ct400247p}{J. Chem. Theory Comput. \textbf{9}, 2959–2973 (2013).}






\end{thebibliography}
\end{document}